\documentclass[11pt,a4paper]{article}

\usepackage{tikz}
\usetikzlibrary{positioning,arrows.meta,calc}
\usepackage{subcaption}
\usepackage{listings}
\usepackage{xcolor}
\usepackage[ruled,vlined]{algorithm2e}
\usepackage[T1]{fontenc}
\usepackage[utf8]{inputenc}
\usepackage[english]{babel}

\usepackage{amsmath,amssymb,amsthm}
\usepackage{bm}

\usepackage{braket}

\usepackage{graphicx}
\usepackage{booktabs}

\usepackage{siunitx}

\usepackage[colorlinks=true,allcolors=blue]{hyperref}

\usepackage[
  left=2.0cm,
  right=2.0cm,
  top=2.5cm,
  bottom=2.5cm
]{geometry}
\newcommand{\op}[1]{\hat{#1}}

\title{\bf Revisiting the  machine-learning density functional for the one-dimensional Hubbard model with random external potential}

\author{Octavio D. R. Salmon,  Minos A. Neto, J. Roberto Viana and Griffith Mendonça\\
\small Departmento de Física do Instituto de Ciências Exatas, \\ Universidade Federal do Amazonas, 3000, Japiim, 69077-000, Manaus-AM, Brazil\\}

\begin{document}
\maketitle

\begin{abstract}
We revisit the machine-learning (ML) approach to the universal density functional $F[\mathbf{n}]$ of the one-dimensional Hubbard model with a site-dependent random potential $\mathbf{v}=\{v_{i}\}$. We generate exact ground-state data via exact diagonalization for a periodic chain with $L=8$ in the paramagnetic sector $(N_\uparrow,N_\downarrow)=(2,2)$, with site electron densities $n_{i} = n_{i\uparrow}=n_{i\downarrow}$. The  resulting density-potential dataset is analyzed. Using principal component analysis of the joint feature space $(\mathbf n,\mathbf v)$, we identify the intrinsic low-dimensional structure of the data.
Then, we restricted the study of the dataset with   
an energy-based filtering criterion to concentrate the data around weakly perturbed energy values with zero potential. A compact one-dimensional convolutional neural network is trained to learn the universal functional considering the lattice periodicity through unilateral wrapping and enforce the lattice symmetries by data augmentation (translations and mirror reflections), achieving near-exact predictions of $F[\mathbf n]$.
Finally, we address the fact that accurate functional values do not necessarily imply accurate functional derivatives. By augmenting training with a variational consistency term that constrains the Euler-Lagrange relation between $\partial F/\partial n_i$ and the gauge-fixed potential we reconstruct the external potentials from automatic differentiation. These results clarify the roles of dataset geometry, symmetry, gauge fixing, and derivative-based constraints in learning physically consistent density functionals.
\end{abstract}

\section{Introduction}
Density Functional Theory (DFT) provides a practical route to the ground-state
properties of interacting quantum many-body systems by replacing the full
many-body wave function with the particle density as the central variable \cite{hohenberg1964,kohn1999}.
In the context of a lattice model with sites $i=1,...,L$, the Hohenberg-Kohn theorems establish that, for fixed interaction and particle
number, (i) the ground-state density $\{n_{i\sigma}\}$ determines the external potential $\{v_{i}\}$ up to an
additive constant and (ii) the ground-state energy follows from a variational
principle. These results imply the existence of a universal functional
$F[\{n_{i}\}$ such that the energy of the ground state  $E_{\mathbf{v}}[\{n_{i\sigma}\}]=F[\{n_{i\sigma}\}]+\sum_{i,\sigma} v_i n_{i\sigma}$ is minimized at the true ground-state density. In practice, however, the exact functional is in general
unknown and must be approximated.\\

Machine learning can approximate $F[\{n_{i\sigma}\}]$ with
high accuracy from data generated by reliable numerical solvers \cite{snyder2012,Polak2025}. In particular, Nelson \textit{et~al.} \cite{nelson2019} demonstrated  that a convolutional neural network trained on exact diagonalization data for the one-dimensional
Hubbard model in a random external potential can learn a highly accurate
approximation to the exact universal functional and reproduce the numerical
manifestations of both HK theorems.\\

The present paper revisits the learning of the Hubbard functional  with emphasis on three aspects that
are crucial for physical reliability: (i) the structure of the dataset induced
by the physical mapping $\,\mathbf v \mapsto \mathbf n(\mathbf v)\,$, including
its intrinsic dimensionality and gauge redundancies; (ii) the role of symmetry
augmentation (translations and mirror reflections) in enforcing lattice
invariances and improving generalization; and (iii) the distinction between
learning functional \emph{values} and learning functional \emph{derivatives}.

While a network may predict the functional almost perfectly, the reconstructed
potential from the predicted functional can still be inaccurate, because standard losses constrain $F[\{n_{i\sigma}\}]$ but not its gradient.\\

To address this, we complement value-based training with an additional
variational consistency constraint inspired by the Euler-Lagrange condition,
together with systematic gauge fixing of the potential  All calculations are implemented in
Python, using \texttt{QuSpin} \cite{Weinberg2024b} for Hamiltonian construction and exact
diagonalization, and standard deep-learning tools for training, such as Convolutional neural networks (CNNs), automatic differentiation, supervised learning, uncertainty–weighted multi–task learning and gradient–based optimization (Adam) \cite{Goodfellow2016,Kendall2018, Yin2026,bishop2006,Ruder2016}.

The remainder of this paper is organized as follows.
In Sec.~2 we introduce the one-dimensional Hubbard model with a site-dependent random potential and discuss the construction of the many-body Hilbert space.
In Sec.~3 we review the lattice formulation of the Hohenberg--Kohn theorems and motivate the use of machine learning for approximating the universal functional. In Sec.~4 we analyze the structure of the dataset through principal component analysis and discuss the effects of energy-based filtering.
Section~5 presents the learning procedure and the convolutional neural network architecture used to approximate the functional. In Sec.~6 we address the reconstruction of the external potential from the functional derivative and introduce a variational consistency constraint. Finally, conclusions are given in Sec.~7.
\section{The one-dimensional Hubbard model with a site-dependent potential}

The Hubbard model for fermions is a simplified lattice model based on the Tight-Binding model to describe the transition between a conducting regime and an insulating one. For this purpose, an onsite energy term is added to consider the interaction between electrons of the same site with opposite spins. Despite its simplicity, research on the Hubbard model continues to have significant impact on the understanding of strongly correlated systems \cite{xu2025}.

The one-dimensional Hamiltonian with L sites with periodic boundary conditions (PBC) can be described as follows
\begin{equation}
\hat{H}
=
- t \sum_{i=1}^{L}
\sum_{\sigma=\uparrow,\downarrow}
\left(
\hat{c}_{i\sigma}^{\dagger}
\hat{c}_{i+1,\sigma}
+
\hat{c}_{i+1,\sigma}^{\dagger}
\hat{c}_{i\sigma}
\right)
+
U \sum_{i=1}^{L}
\hat{n}_{i\uparrow}
\hat{n}_{i\downarrow}
+
\sum_{i=1}^{L}
v_i
\hat{n}_i ,
\label{eq:Hubbard}
\end{equation}

where $\hat{c}_{i\sigma}^{\dagger}$ ($\hat{c}_{i\sigma}$) creates (annihilates)
a fermion with spin $\sigma=\uparrow,\downarrow$ at site $i$.
The first term describes nearest-neighbor hopping with amplitude $t$,
allowing particles to move along the lattice.
The second term accounts for the on-site interaction with strength $U$. The last term represents an external site-dependent potential $v_i$ coupled
to the local density
\begin{equation}
\hat{n}_i = \hat{n}_{i\uparrow} + \hat{n}_{i\downarrow},
\qquad
\hat{n}_{i\sigma} = \hat{c}_{i\sigma}^{\dagger}\hat{c}_{i\sigma}.
\end{equation}
The PBC are assumed, such that $\hat{c}_{L+1,\sigma} \equiv \hat{c}_{1,\sigma}$.

The clean one-dimensional Hubbard model is exactly solvable via the Bethe ansatz and exhibits no Mott transition at half filling for finite $U>0$ \cite{liebwu1968,essler2009}. Introducing random on-site potentials ($v_i$) destroys integrability, so in order to solve it generally requires numerical approaches.\\

For the one-dimensional Hubbard model defined on a lattice with \(L\) sites, each site \(i\) supports four local electronic states,
\begin{equation}
\{|0\rangle,|\uparrow\rangle,|\downarrow\rangle,|\uparrow\downarrow\rangle\}.
\end{equation}
The total Hilbert space is therefore given by the tensor product
\begin{equation}
\mathcal{H}
=
\bigotimes_{i=1}^{L} \mathcal{H}_i,
\qquad
\dim(\mathcal{H}) = 4^{L}.
\end{equation}

In the absence of spin-flip terms, the Hubbard Hamiltonian conserves separately the number of spin-up and spin-down particles,$\hat N_\uparrow $ and $\hat N_\downarrow $,
so that the Hilbert space decomposes into invariant sectors labeled by
\((N_\uparrow,N_\downarrow)\),
\begin{equation}
\mathcal{H}
=
\bigoplus_{N_\uparrow=0}^{L}
\bigoplus_{N_\downarrow=0}^{L}
\mathcal{H}_{(N_\uparrow,N_\downarrow)},
\end{equation}
where
\begin{equation}
\dim \mathcal{H}_{(N_\uparrow,N_\downarrow)}
=
\binom{L}{N_\uparrow}\binom{L}{N_\downarrow}.
\end{equation}

Equivalently, one may label sectors by the conserved total particle number
\(N=N_\uparrow+N_\downarrow\) and the conserved magnetization
\begin{equation}
S_z=\frac{1}{2}(N_\uparrow-N_\downarrow).
\end{equation}
In this representation, the Hilbert space admits the hierarchical decomposition
\begin{equation}
\mathcal{H}
=
\bigoplus_{N}
\bigoplus_{S_z}
\mathcal{H}_{N,S_z},
\qquad
\mathcal{H}_{N,S_z}
\equiv
\mathcal{H}_{(N_\uparrow,N_\downarrow)}.
\end{equation}

Of particular interest is the sector of vanishing magnetization, $S_z=0$, which corresponds to equal numbers of spin--up and spin--down
particles,
\(
N_\uparrow = N_\downarrow.
\)
In this case, the Hilbert space decomposes as a direct sum of all
\((N_\uparrow,N_\downarrow)\) sectors satisfying this condition,
\begin{equation}
\mathcal{H}_{S_z=0}
=
\bigoplus_{m=0}^{L}
\mathcal{H}_{(m,m)}.
\end{equation}

For a chain of length \(L=8\) (with periodic boundary conditions),
this decomposition reads explicitly
\begin{align}
\mathcal{H}_{S_z=0}
&=
\bigoplus_{m=0}^{8}\mathcal{H}_{(m,m)}
\nonumber\\
&=
\mathcal{H}_{(0,0)}\oplus \mathcal{H}_{(1,1)}\oplus \mathcal{H}_{(2,2)}\oplus \mathcal{H}_{(3,3)}
\oplus \mathcal{H}_{(4,4)}\oplus \mathcal{H}_{(5,5)}\oplus \mathcal{H}_{(6,6)}\oplus
\mathcal{H}_{(7,7)}\oplus \mathcal{H}_{(8,8)}.
\end{align}

In the present paper we restrict the system to paramgnetic sector $N_{\uparrow} = N_{\downarrow}$. Accordingly, the many-body Hilbert space is constructed in the occupation-number (Fock) basis, formed by eigenstates of the local number operators $\hat{n}_{i\sigma}$. So, the basis state can be written as
\begin{equation}
\lvert \alpha \rangle
=
\bigotimes_{i=1}^{L}
\lvert n_{i\uparrow}, n_{i\downarrow} \rangle,
\qquad
n_{i\sigma} \in \{0,1\},
\end{equation}
where $n_{i\sigma}$ denotes the occupation of site $i$ by a fermion
with spin $\sigma$. Therefore, the restricted Hilbert space is given by
\begin{equation}
\mathcal{H}_{N_{\uparrow},N_{\downarrow}}
=
\mathrm{span}
\left\{
\lvert \alpha \rangle :
\sum_i n_{i\uparrow}=N_{\uparrow},
\;
\sum_i n_{i\downarrow}=N_{\downarrow}
\right\}.
\end{equation}

Here, $\mathrm{span}\{\cdots\}$ denotes the linear subspace generated by all occupation--number basis states satisfying the specified particle--number constraints, i.e., the set of all complex linear combinations of such basis
states.

The \texttt{QuSpin} library provides a transparent and modular framework for
constructing many-body Hamiltonians in second quantization. Rather than explicitly building matrix elements by hand, the Hamiltonian is specified as a sum of operator terms, each defined by its algebraic structure and coupling constants  \cite{Weinberg2024b}.

We employed \texttt{QuSpin}  not only to construct the Hubbard Hamiltonian as a sparse matrix, but also to perform its diagonalization and to compute ground-state observables, including the local site occupations
$n_i=\langle \op{n}_{i\uparrow}+\op{n}_{i\downarrow}\rangle$, 
\footnote{It is important to clarify that in the occupation-number (Fock) basis, the symbols $n_{i\sigma}$ denote the eigenvalues of the local number operators  $\op{n}_{i\sigma}=\op{c}^\dagger_{i\sigma}\op{c}_{i\sigma}$ and therefore take only the discrete values $n_{i\sigma}\in\{0,1\}$. In contrast, when discussing density functional theory, the same symbol $n_{i\sigma}$ is often used to denote the local electronic density, defined as
the ground-state expectation value $n_{i\sigma}
\equiv \langle \mathrm{GS} | \op{n}_{i\sigma} | \mathrm{GS} \rangle$, where $0 \le n_{i\sigma} \le 1.$}, which is specifically  $\langle \mathrm{GS} | \op{n}_{i\uparrow} | \mathrm{GS} \rangle + \langle \mathrm{GS} | \op{n}_{i\downarrow} | \mathrm{GS} \rangle  = n_{i\uparrow} + n_{i\downarrow}$.


\section{Hohenberg--Kohn theorems and the role of machine learning}

Density Functional Theory is founded on two fundamental results known as the
Hohenberg--Kohn (HK) theorems. Although originally formulated for interacting
electrons in the continuum, these theorems admit a natural analogue for lattice
models such as the Hubbard Hamiltonian.

The first HK theorem establishes a one-to-one correspondence between the ground-state density and the external potential, up to an additive constant. In the
context of the Hubbard model, the role of the electronic density is played by
the set of local site occupations $\{n_{i\sigma}\}$, and the theorem states that
the ground-state density uniquely determines the external potential $\{v_i\}$, apart from an irrelevant constant shift. Therefore, all ground-state observables are functionals of the density alone.

This result implies the existence of a \emph{universal functional}
\begin{equation}
F[\{n_{i\sigma}\}]
=
\langle \mathrm{GS} | \op{T} + \op{V}_{int} | \mathrm{GS} \rangle,
\end{equation}
which is independent of the external potential and depends only on the kinetic $\op{T}$
and interaction terms $\op{V}_{int}$ of the Hamiltonian.

The second HK theorem provides a variational principle for the ground-state energy. For any trial density $\{n_{i\sigma}\}$ compatible with the particle
number, the total energy functional
\begin{equation}
E[\{n_{i\sigma}\}]
=
F[\{n_{i\sigma}\}]
+
\sum_{i,\sigma} v_i\,n_{i\sigma}
\label{eq:identidade}
\end{equation}
satisfies $E[\{n_{i\sigma}\}] \ge E_{\mathrm{GS}} $, 
with equality if and only if the density coincides with the true ground-state
density. Consequently, the ground state can be obtained by minimizing the energy functional with respect to the density.

While the HK theorems guarantee the existence and properties of the exact functional, they do not provide its explicit form. So, in general, the functional must be approximated. Machine learning offers a powerful and flexible strategy to address this limitation.

By training a neural network on exact ground-state data obtained from 
diagonalization of the Hubbard Hamiltonian for many different external
potentials, machine learning can construct an explicit approximation to the
universal functional $F[\{n_{i\sigma}\}]$. In this approach, the neural network learns the mapping $
\{n_{i\sigma}\}
\;\longrightarrow\;
F[\{n_{i\sigma}\}],
$
directly from data, without assuming a specific analytical form.

Within this framework, the first HK theorem is tested numerically by verifying that the learned functional establishes a unique correspondence between ground-
state densities and external potentials.

In order to generate the dataset the Hubbard Hamiltonian is evaluated for a large number of different external potentials $\{v_i\}$. This requires repeatedly updating the Hamiltonian and computing its ground-state properties, a task for which \texttt{QuSpin} is particularly well suited. 

It is important to clarify that the density functional $F$ is expressed in terms of the spin-resolved local densities $n_{i\uparrow}$ and $n_{i\downarrow}$, since $n_{i} = n_{i\uparrow} + n_{i\downarrow} $.
However, we are restricted in this paper to the paramagnetic sector, $N_\uparrow = N_\downarrow$, where the ground-state densities satisfy
\begin{equation}
n_{i\uparrow} = n_{i\downarrow}
\qquad \text{for all } i .
\end{equation}
As a consequence, the density functional depends only on of one of the two spin densities, so $F[\{n_{i\sigma}\}] = F[\{n_{i\uparrow}\}] = F[\{n_{i\downarrow}\}]$.
For notational simplicity, we therefore redefine
\begin{equation}
n_i \equiv n_{i\uparrow} = n_{i\downarrow},
\end{equation}
and write the universal functional as
\begin{equation}
F = F[\{n_i\}] .
\end{equation}

With this convention, the external potential contribution to the Hubbard
Hamiltonian,
\begin{equation}
\hat H_v = \sum_{i,\sigma} v_i \hat n_{i\sigma},
\end{equation}
leads to the following expression after calculating the ground-sate mean 
\begin{equation}
\langle \hat H_v \rangle
=
2 \sum_i v_i n_i ,
\end{equation}
where the factor of two accounts for the spin degeneracy.
All subsequent expressions are written in terms of the single-site density $n_i$ defined above.



\section{Analysis of the Dataset}

The dataset employed in this work is generated from the one-dimensional Hubbard model with $N=L$ sites in PBC subject to a site-dependent random potential $\{v_{i}\}$. 

For each disorder realization, the on-site potentials $v_i$ are drawn independently from a uniform distribution $v_i \in [-W, W]$, with the disorder amplitude $W$ taking discrete values in the interval $W \in [0.005t,\,2.5t]$. These discrete values were selected as $W_k = W_{\min} + k\,\Delta W$ with
$W_{\min}=0.005t$, $W_{\max}=2.5t$, and $N_W=61$.
For each value of $W$, a large number of realizations were generated and the ground-state energy $E$ and density profile $\mathbf{n} = (n_{1},...,n_{L}) = $ were computed exactly. 
Following the parameters used in Ref.~\cite{nelson2019}, we choose $L=8$ in the paramagnetic sector $N_{\uparrow} = N_{\downarrow}=2$. The total number of samples generates in the dataset is $N_{tot} = 600k$, where for each one, $\{n_{i}\}$, $\{v_{i}\}$, $W$, $E$ and $F$ were stored. Furthermore, the whole dataset contains the same number of samples for each value of $W$.

We denote $E^{homo}$ and ${\mathbf{n}}^{\mathrm{homo}}$ as the ground-state energy and the density vector corresponding to the 
homogeneous system with the same parameters $(t,U,N)$ in which  $v_{i}=0$ for all sites. Accordingly,  the norm given by 

\begin{equation}
\left|\mathbf{n}-\mathbf{n}^{\mathrm{homo}}
\right|= \sqrt{
\sum_{\sigma = \uparrow}^{\downarrow}\sum_{i=1}^{L}
\bigl(n_{i\sigma} - n_{i\sigma}^{\mathrm{homo}} \bigr)^2} =\sqrt{2\sum_{i=1}^{L} \bigl(n_i - \tfrac{1}{4}\bigr)^2}
\label{eq:dist_density}
\end{equation}

is the distance between the density vector and the homogeneous density. The calculations of the homogeneous system give $n_{i\sigma}^{homo}=1/4$ and $E^{homo}=-5.9517 \approx -6$, which is in units of the hopping energy $t$, which we set as 1.

\begin{figure}[h]
    \centering
    
    \begin{subfigure}{0.45\textwidth}
        \centering
        \includegraphics[width=\linewidth]{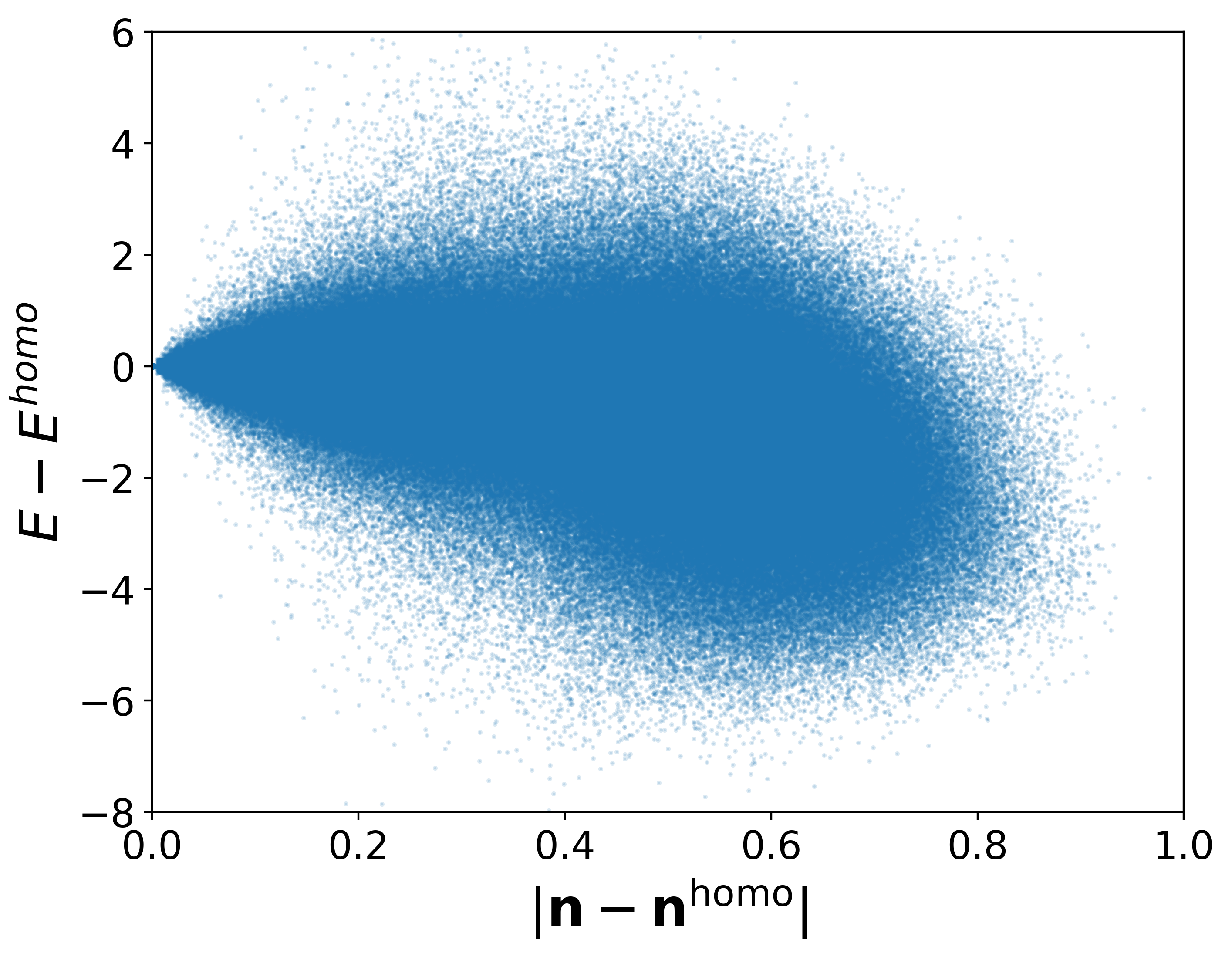}
        \caption{}
        \label{fig:a}
    \end{subfigure}
    \begin{subfigure}{0.45\textwidth}
        \centering
        \includegraphics[width=\linewidth]{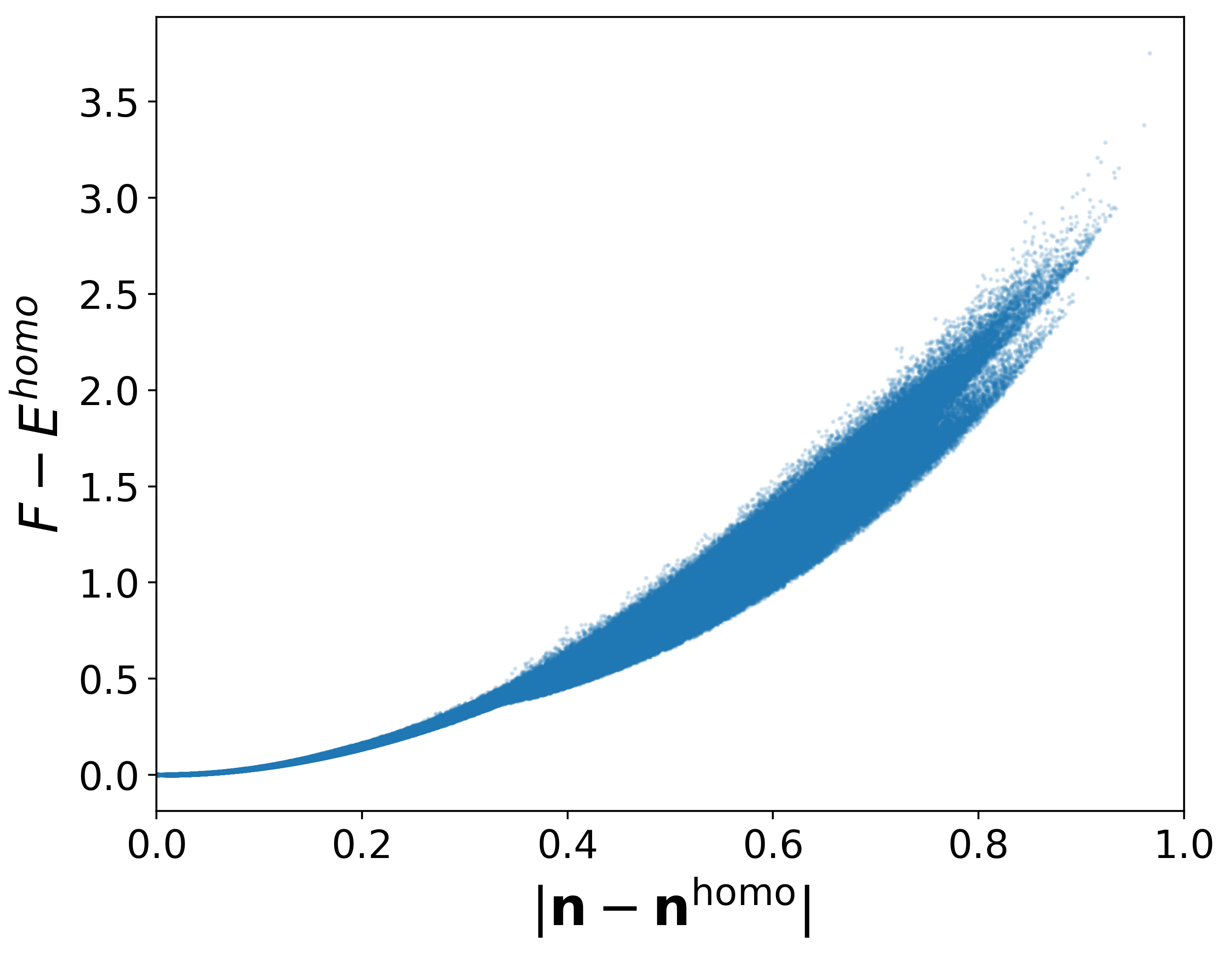}
        \caption{}
        \label{fig:b}
    \end{subfigure}
      \caption{In (a) is shown the ground-state energy $E$ relative to the ground-sate energy of the homogeneous system $E^{homo}$. In (b) the universal functional $F$ relative to $E^{homo}$.} 
    \label{fig:fignonfiltered}
\end{figure}

In Fig. \ref{fig:fignonfiltered} are plotted $E-E^{homo}$, $F_E^{homo}$ as functions of the norm ${\mathbf{n}}^{\mathrm{homo}}$. Note that the ground energy $E$ exhibits a high dependence of the random potential, whereas $F$ shows its universal character depending only on the electron density for given values of $U$ and $t$. Nevertheless, for greater values of the norm, the functional $F$ shows non-monotonic behavior. So, according to Fig. \ref{fig:fignonfiltered} (b) , $F$ is monotonic for values close to $E^{homo}$.

Although the joint feature vector $(n_\uparrow, v) \in \mathbb{R}^{2L} $ lives in a $2L$-dimensional space, it does not fill this space arbitrarily. Instead, it lies on the graph of the physical mapping
\begin{equation}
\mathbf{v} \longmapsto \mathbf{n}_\uparrow(\mathbf{v}) = \{n_{i}(\mathbf{v})\},
\end{equation}
which defines a manifold embedded in $\mathbb{R}^{2L}$.

Principal Component Analysis (PCA) may provide a  way to characterize the intrinsic dimensionality and geometry of this manifold. By diagonalizing the covariance matrix of the joint dataset, PCA identifies the dominant directions of variance and thus reveals how the graph of $n_\uparrow(v)$ is embedded in feature space.

In order to perform this analysis, we consider, for each sample realization, the joint feature vector
\begin{equation}
x^{(a)} =
\left(
n_{1}^{(a)}, \dots, n_{8}^{(a)},
v_1^{(a)}, \dots, v_8^{(a)}
\right)
\in \mathbb{R}^{16}.
\end{equation}

The complete dataset defines the matrix $ X \in \mathbb{R}^{N_{\mathrm{samples}} \times 16}$. 

Before applying PCA, each column is standardized:
\begin{equation}
\tilde X_{ij}
=
\frac{X_{ij} - \mu_j}{\sigma_j},
\end{equation}
where $\mu_j$ and $\sigma_j$ are the empirical mean and standard deviation of feature $j$.

PCA diagonalizes the empirical covariance matrix
\begin{equation}
C = \frac{1}{N} \tilde X^T \tilde X.
\end{equation}

The principal components are obtained from the eigenvalue problem
\begin{equation}
C u_k = \lambda_k u_k,
\end{equation}
where $\lambda_k$ are the eigenvalues and $u_k$ the orthonormal eigenvectors.

The variance explained by component $k$ is
\begin{equation}
\mathrm{Var}_k = \frac{\lambda_k}{\sum_j \lambda_j}.
\end{equation}

\begin{figure}
    \centering
    \includegraphics[width=0.6\linewidth]{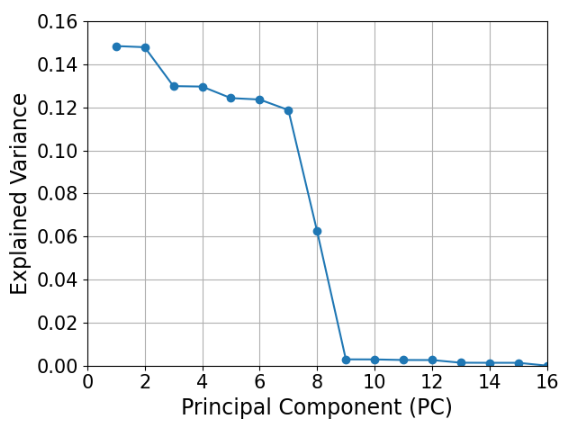}
    \caption{Explained Variance versus principal components for the feature space $(n_{1},...,n_{L},v_{1},...,v_{L})$, for $L=8$ and $(2,2)$ sector. The figure shows that relevant components are the first eight ones. }
    \label{fig:pca1}
\end{figure}

Fig. \ref{fig:pca1} displays the variance explained. The spectrum exhibits a clear separation Components PC1-PC8 carry significant variance, whereas PC9-PC16 have negligible values.

This structure reflects the intrinsic dimensionality of the dataset. However, it is important to analyse PC separately.

To quantify the relative contribution of density and potential degrees of freedom in each principal component, we decompose every eigenvector of the PCA into two blocks.

Let $u_k \in \mathbb{R}^{16}$ denote the $k$-th principal component (eigenvector of the covariance matrix).
We partition it as
\begin{equation}
u_k = \big( u_k^{(n)},\, u_k^{(v)} \big),
\end{equation}
where
\[
u_k^{(n)} \in \mathbb{R}^8, 
\qquad
u_k^{(v)} \in \mathbb{R}^8.
\]

We define the block norms
\begin{equation}
\|u_k^{(n)}\| = 
\left( \sum_{i=1}^{8} (u_{k,i}^{(n)})^2 \right)^{1/2},
\qquad
\|u_k^{(v)}\| = 
\left( \sum_{i=1}^{8} (u_{k,i}^{(v)})^2 \right)^{1/2}.
\end{equation}

Since the principal components are normalized,
\begin{equation}
\|u_k\|^2
=
\|u_k^{(n)}\|^2
+
\|u_k^{(v)}\|^2
=
1.
\end{equation}

To measure the relative weight of the potential sector in each principal component, we introduce the ratio
\begin{equation}
R_k
=
\frac{\|u_k^{(v)}\|}{\|u_k^{(n)}\|}.
\end{equation}

This quantity has a clear geometric meaning.  $R_k \approx 1$ indicates that density and potential contributions are comparable. $R_k \gg 1$ indicates that the component is dominated by the potential block.  $R_k \ll 1$ indicates dominance of the density block. 

\begin{figure}
    \centering
    \includegraphics[width=0.6\linewidth]{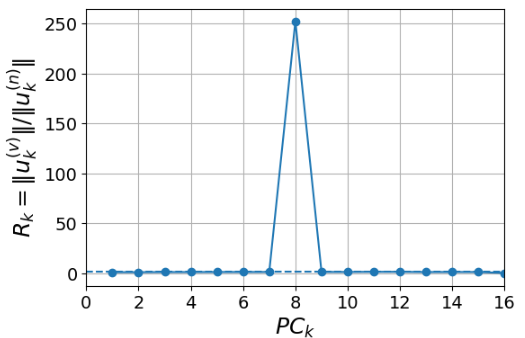}
    \caption{ Ratio $R_k = \|u_k^{(v)}\| / \|u_k^{(n)}\|$ for each principal component of the joint density–potential dataset. Here $u_k^{(n)}$ and $u_k^{(v)}$ denote the density and potential blocks of the $k$-th PCA eigenvector.
Components PC1-PC7 exhibit comparable density and potential contributions ($R_k \sim 1$), characteristic of genuine density-response directions. The pronounced peak at PC8 reveals a component almost entirely confined to the potential subspace.
}
 \label{fig:pca2}
\end{figure}

In Fig.\ref{fig:pca2} is shown the ratio between the potential and density block norms of each principal component,
$R_k = \|u_k^{(v)}\| / \|u_k^{(n)}\|$, obtained from the PCA of the joint feature vector $(n_\uparrow, v)$.
Components PC1-PC7 exhibit comparable density and potential contributions ($R_k \sim 1$), indicating physical response directions. The pronounced peak at PC8 reveals a mode almost entirely confined to the potential subspace.
For clarification, for the first six principal components we obtain
\begin{equation}
\|u_k^{(n)}\| \sim 0.7,
\qquad
\|u_k^{(v)}\| \sim 0.7,
\qquad
R_k \approx 1.
\end{equation}

This implies that PC1-PC6 correspond to directions in the joint space where density and potential vary simultaneously. Geometrically, these components are tangent directions to the manifold $ \mathcal{M} = \{ (n_\uparrow(v),\, v) \}$, 
reflecting genuine physical response directions.

In contrast, PC8 exhibits
\begin{equation}
\|u_8^{(n)}\| \approx 0,
\qquad
\|u_8^{(v)}\| \approx 1,
\qquad
R_8 \gg 1,
\end{equation}
leading to a pronounced peak in $R_k$.

Inspection of the potential block of PC8 reveals that
\begin{equation}
u_8^{(v)} \propto (1,1,1,1,1,1,1,1),
\end{equation}
i.e., it corresponds to a uniform shift of all on-site potentials. Thus, this principal component corresponds to a direction in which all on-site potentials vary simultaneously by the same amount. Such variations are of the form

\begin{equation}
v_i \rightarrow v_i + c,
\end{equation}
so the Hamiltonian transforms as
\begin{equation}
\hat H \rightarrow \hat H + c \hat N.
\end{equation}

Since calculations are performed in a fixed-$N$ sector,\(\hat N |\Psi_0\rangle = N |\Psi_0\rangle\), this transformation produces only an additive shift of the ground-state energy and leaves the density invariant:
\begin{equation}
n_{i} = n_{i\uparrow}(\mathbf{v}+\mathbf{c}) = n_{i\uparrow}(\mathbf{v}).
\end{equation}

Therefore, PC8 represents a gauge degree of freedom of the Hamiltonian rather than a genuine physical response mode, since $\delta \mathbf v \propto \mathbf 1  \Longrightarrow \delta \mathbf n = 0$. This analysis demonstrates that purely data-driven dimensionality reduction can detect gauge redundancies automatically, but interpreting such components requires knowledge of the underlying physical symmetries.\\

In addition, in the paramagnetic sector considered here ($N_\uparrow = 2$ for $L=8$), the spin-resolved densities satisfy the exact constraint
\begin{equation}
\sum_{i=1}^{8} n_{i} = 2.
\end{equation}

This linear relation confines all density vectors to the hyperplane
\begin{equation}
H = \left\{ x \in \mathbb{R}^8 \mid \sum_i x_i = 2 \right\},
\end{equation}
which has dimension $\dim(H) = 8 - 1 = 7$.

This explains the numerical observation that only seven principal components exhibit significant density contributions, while additional components correspond either to gauge redundancies or to negligible variance directions.\\

More generally, for a system with $L$ sites and fixed particle number, the intrinsic dimensionality of the density manifold is $L-1$. While the reduction of dimensionality from $L$ to $L-1$ follows trivially from particle-number conservation, the relevance of the PCA analysis lies in demonstrating that this constraint is automatically encoded in the empirical covariance structure of the dataset.\\

Therefore, the PCA reveals that no significant variance directions exist in the potential subspace beyond the trivial gauge mode, this analysis alone does not establish functional independence of $F$ from $v$. It must instead be assessed by comparing the predictive accuracy of models trained with density-only inputs against models that include both density and potential. If the inclusion of $v$ does not reduce the generalization error on unseen data, this provides empirical evidence that $F$ depends only on $\{n_{i}\}$.

\subsection{Restricting Energy Range}
Following the procedure introduced in Ref.~\cite{nelson2019}, an energy-based filtering criterion is applied in order to restrict the dataset to weakly excited configurations. This is done "in order to prevent the dataset from having large fluctuations in the total energy". So, only samples satisfying the filter
\begin{equation}
\label{eq:energy_filter}
F - E^{homo} < 0.15t
\end{equation}
are retained. Consequently, the dataset of $600k$ samples is reduced to $N_{filt}=178109$ samples after applying this filter. 

\begin{figure}[h]
    \centering
    
    \begin{subfigure}{0.45\textwidth}
        \centering
        \includegraphics[width=\linewidth]{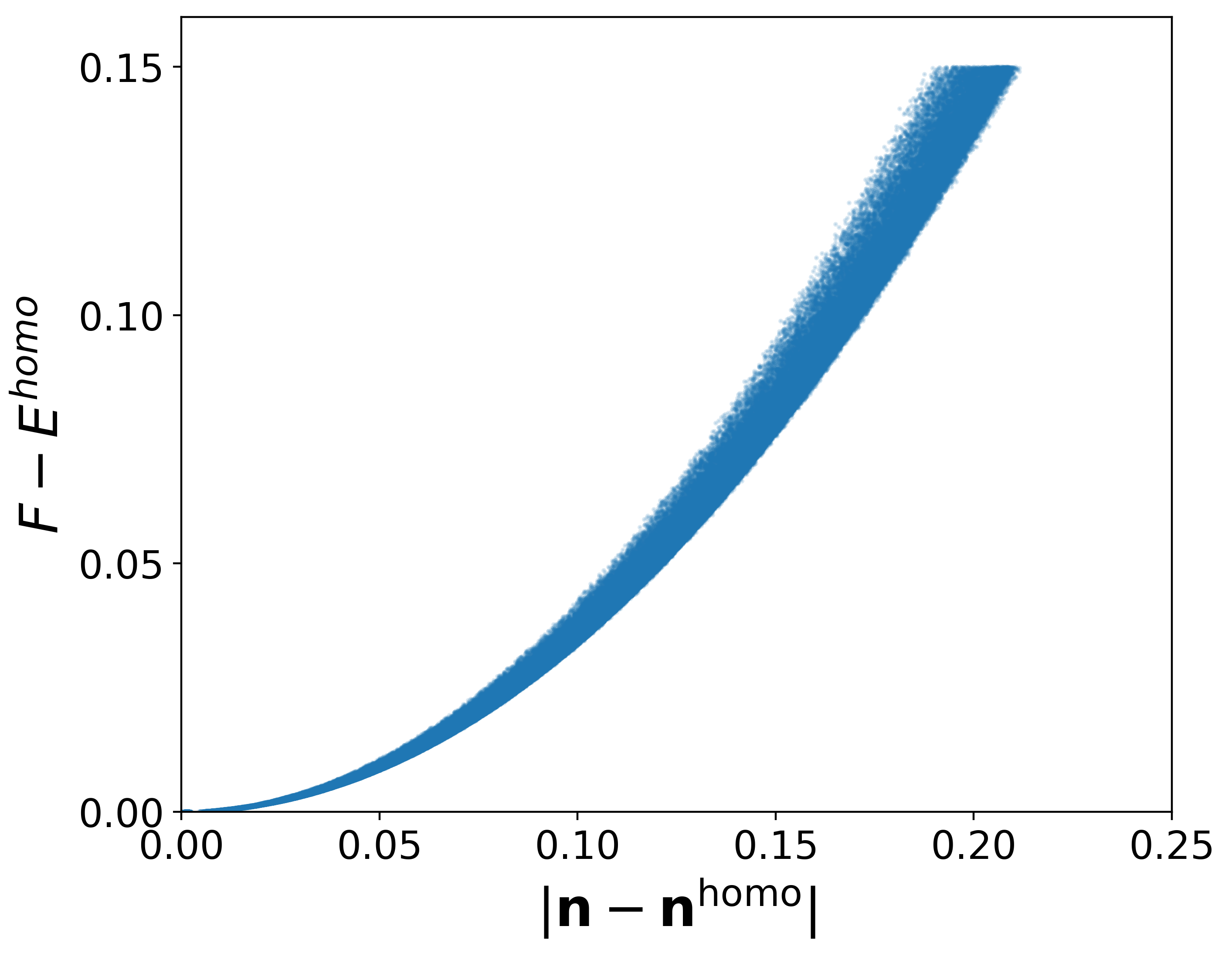}
        \caption{}
        \label{fig:a}
    \end{subfigure}
    \begin{subfigure}{0.45\textwidth}
        \centering
        \includegraphics[width=\linewidth]{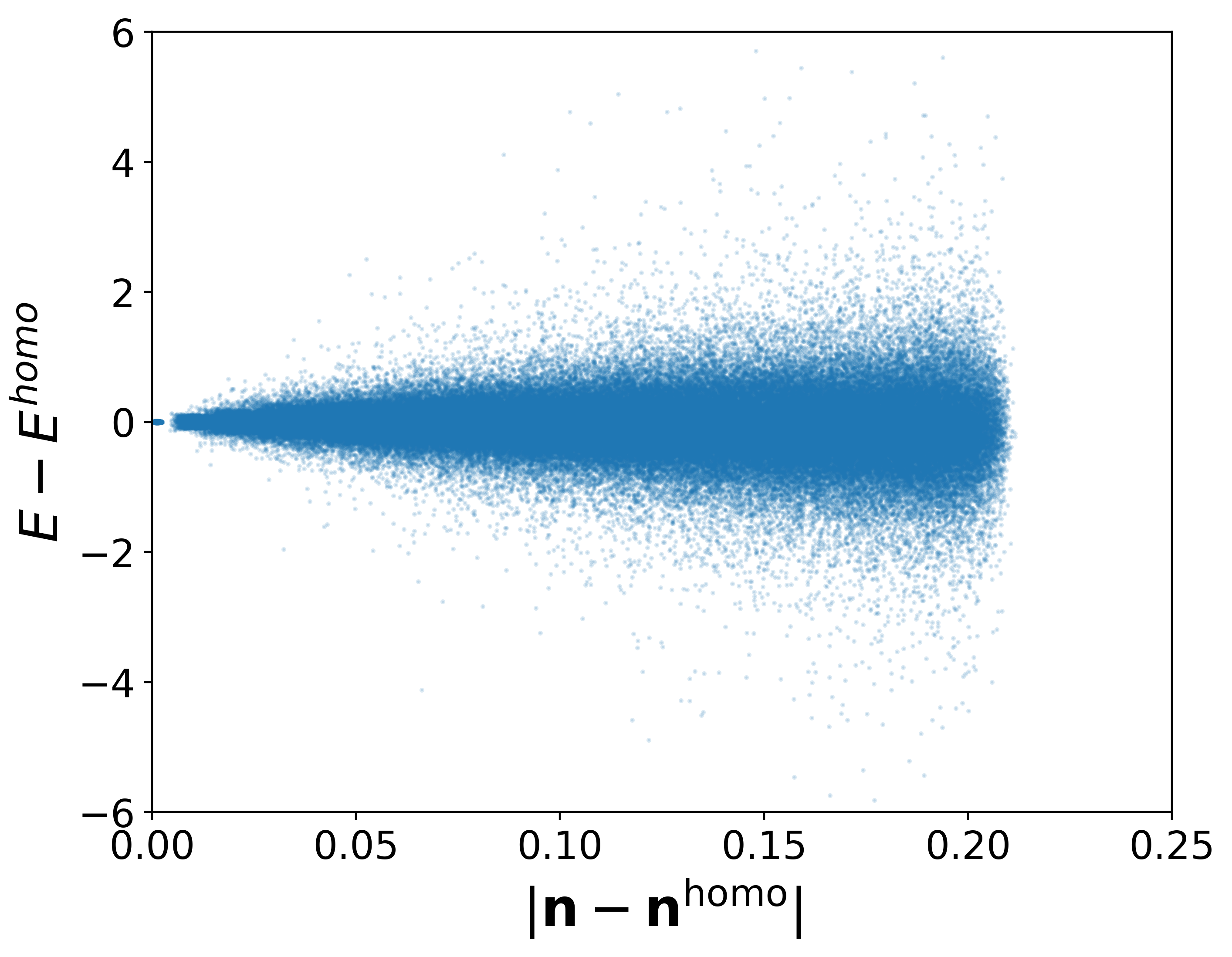}
        \caption{}
        \label{fig:b}
    \end{subfigure}
       \caption{The figures are plotted with a dataset containing filtered samples satisfying the condition $F-E^{homo} < 0.15t$. In (a) the universal functional $F$ relative to $E^{homo}$. In (b) is shown the ground-state energy $E$ relative to the ground-sate energy of the homogeneous system $E^{homo}$. }
    \label{fig:figfiltered}
\end{figure}

Fig. \ref{fig:figfiltered} shows in panel (a) and (b) $F$ and $E$ relative to $E^{homo}$, respectively, as functions of the density norm. Note that $F$ is closer to a monotonic function of the norm, and the ground state energy $E$ exhibits much less fluctuations around $E^{homo}$, but reflects strong influence of the random potential. On the other hand, $F$ shows no direct dependence on the potential due to its universal character. Although the authors in Ref.~\cite{nelson2019} seem to mean that they applied the filtering condition $E-E^{homo}< 0.15t$, the dataset of the present work shows that this is not the case. Furthermore, it can be observed in Fig.1 of Ref.~\cite{nelson2019} a plot of $E - E^{\mathrm{homo}}$ versus the density norm, which is just the same resulting figure as Fig. \ref{fig:figfiltered}(a) in the present paper. So, the vertical axis label in Fig.1 of Ref.~\cite{nelson2019} must be $F - E^{homo} < 0.15t$ and not $E-E^{homo}$.\\

\begin{figure}[h]
    \centering
    
    \begin{subfigure}{0.45\textwidth}
        \centering
        \includegraphics[width=\linewidth]{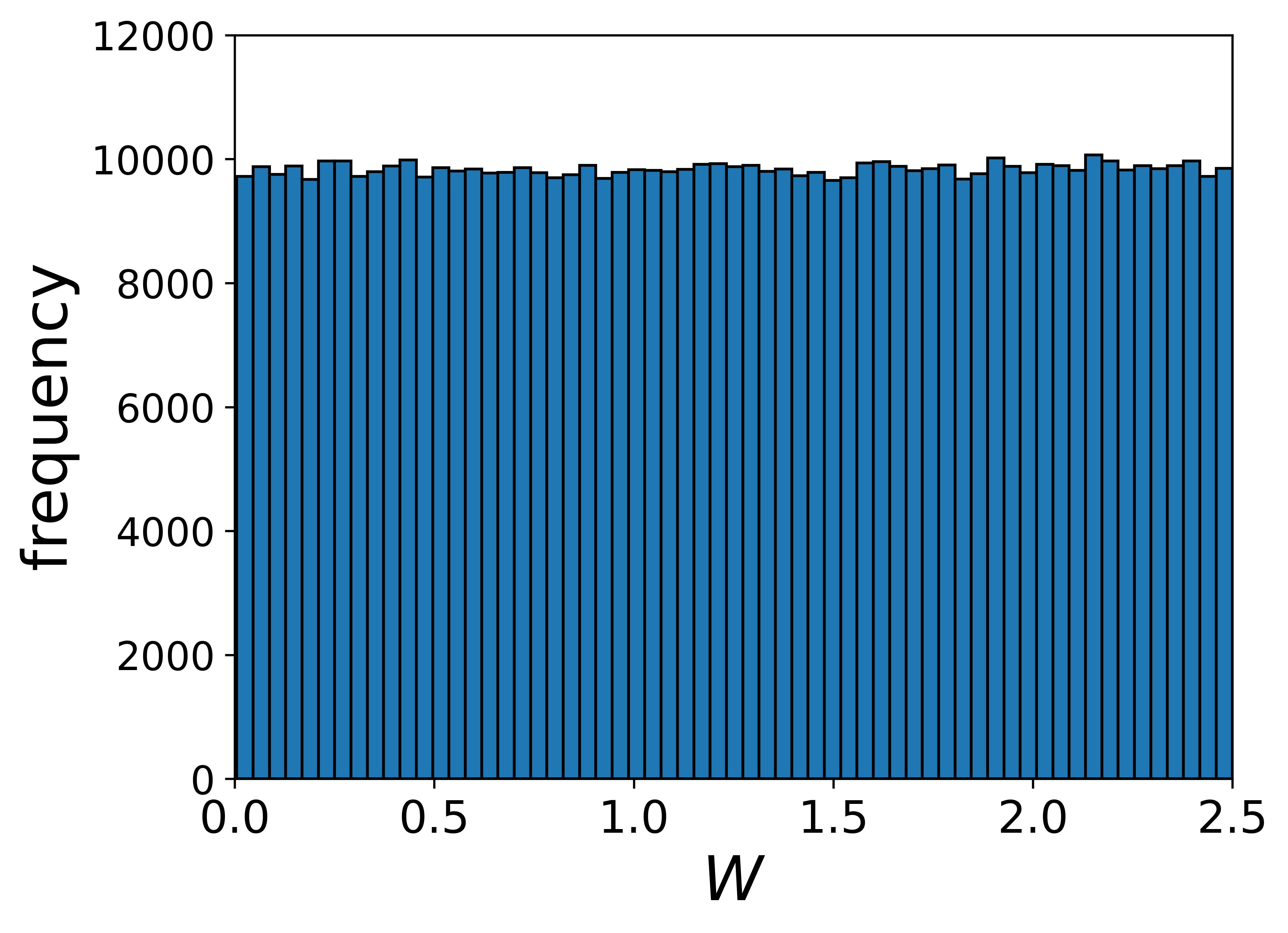}
        \caption{}
        \label{fig:a}
    \end{subfigure}
    \begin{subfigure}{0.45\textwidth}
        \centering
        \includegraphics[width=\linewidth]{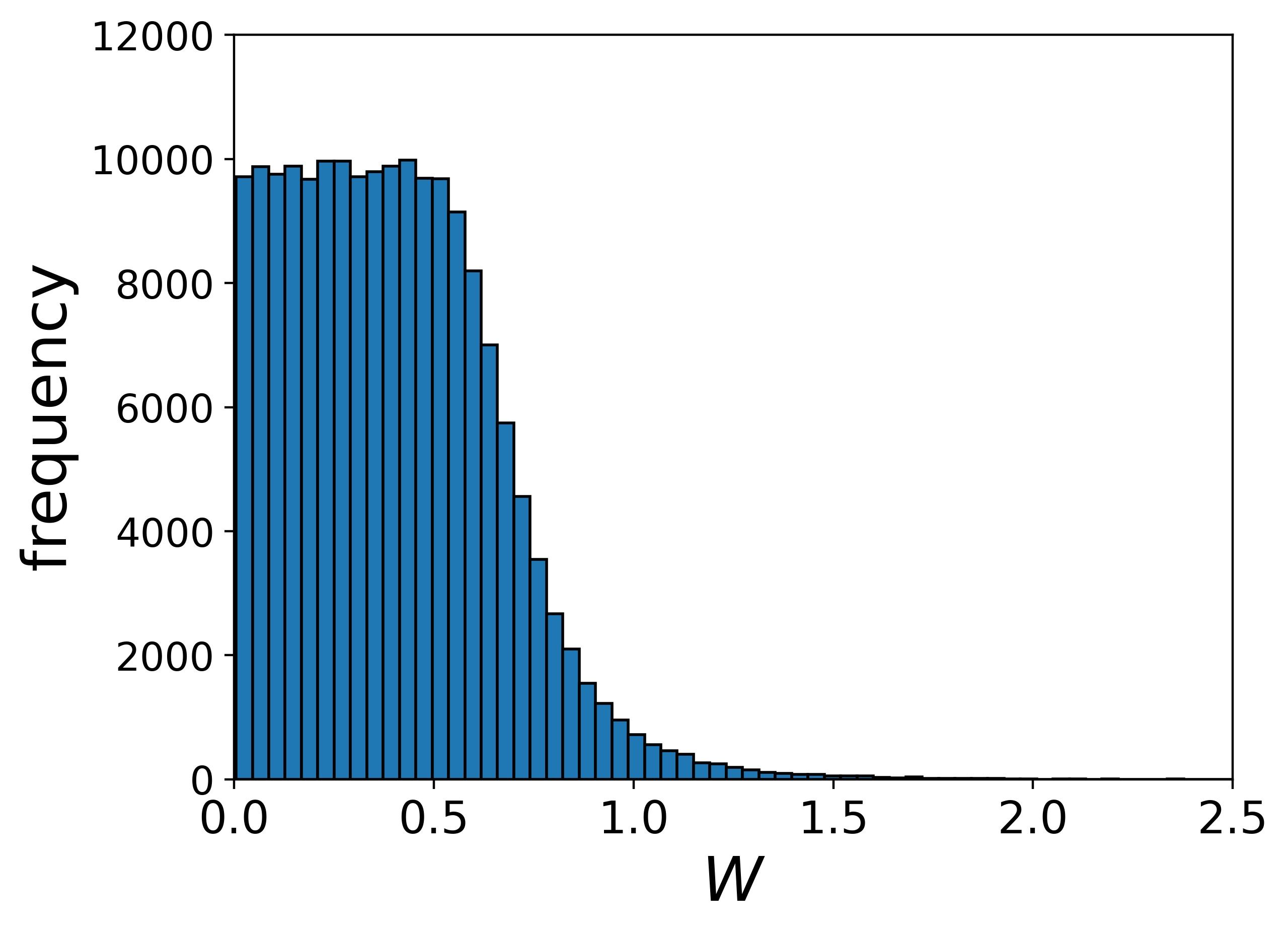}
        \caption{}
        \label{fig:b}
    \end{subfigure}
    \caption{In (a) distribution of the samples of the dataset by the random strength $W$ without filtering. In (b) is exhibited the distribution of the samples after filtering with the condition $F-E^{homo} < 0.15t$.  }
    
    \label{fig:histo1}
\end{figure}

In Fig.\ref{fig:histo1} is shown the distribution of the samples according to the values of $W$. The original dataset was generated with equal numbers of samples per each value of $W$ (see Fig.\ref{fig:histo1}(a)). After filtering with the $F - E^{homo} < 0.15t$ restriction, the frequency of samples according to the value of $W$ changes as shown in Fig.\ref{fig:histo1}(b). So, the number of samples with $v_{i}$ in $[-W,W]$ for $W > 1.0$ is highly reduced. 

\begin{figure}[h]
    \centering
    
    \begin{subfigure}{0.45\textwidth}
        \centering
        \includegraphics[width=\linewidth]{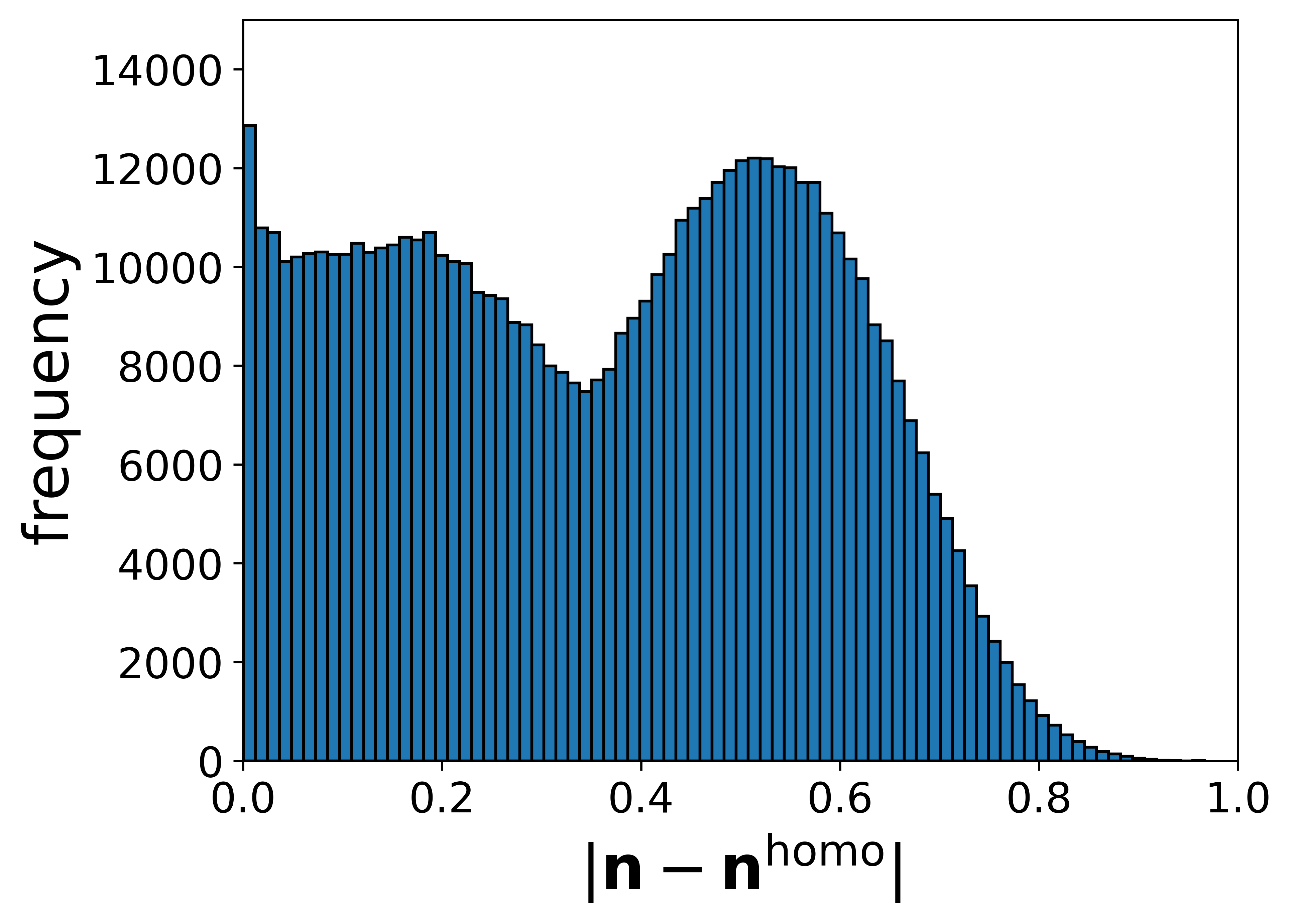}
        \caption{}
        \label{fig:a}
    \end{subfigure}
    \begin{subfigure}{0.45\textwidth}
        \centering
        \includegraphics[width=\linewidth]{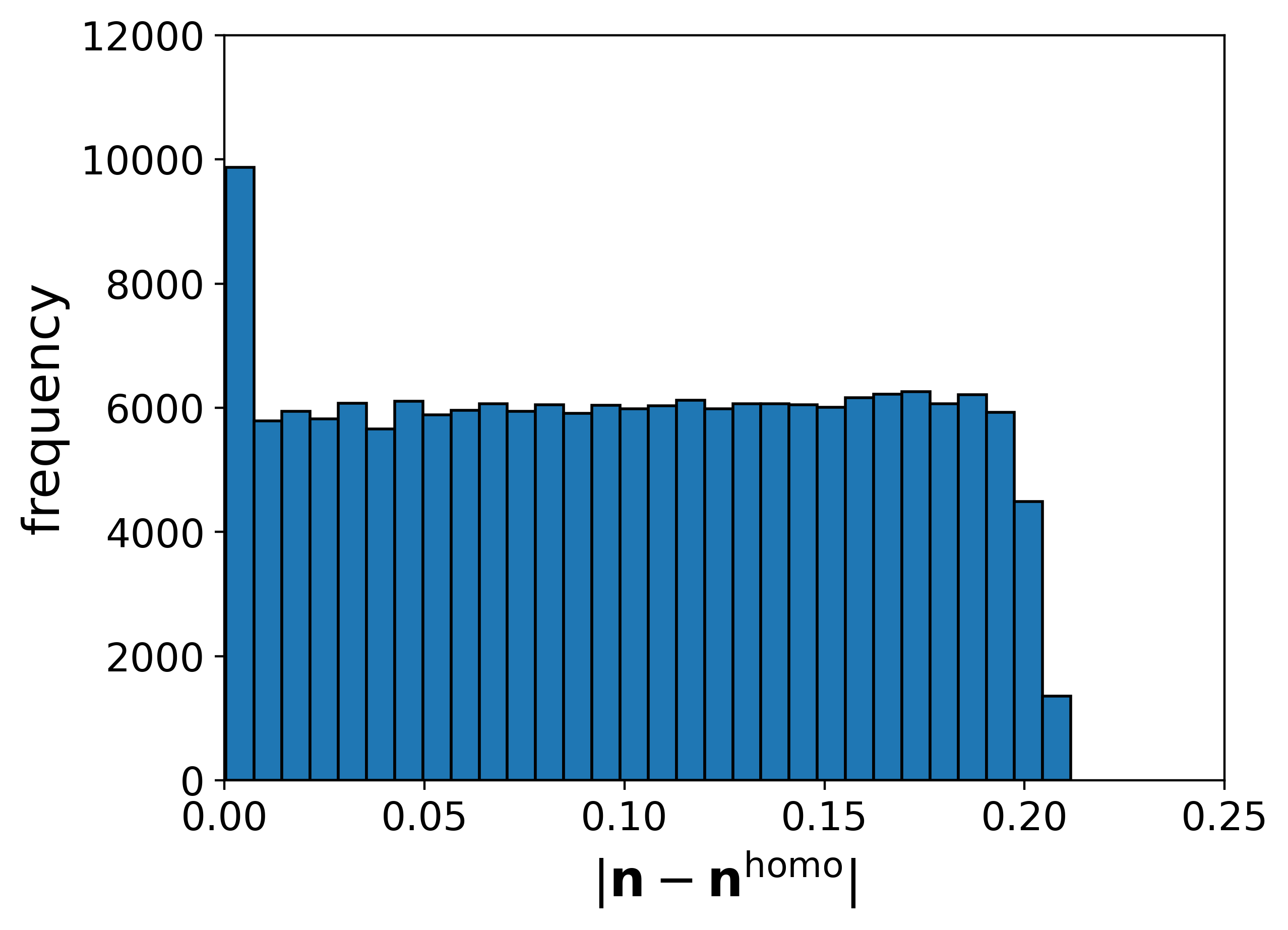}
        \caption{}
        \label{fig:b}
    \end{subfigure}
    \caption{In (a) distribution of density norm for the original dataset. In (b) is exhibited the same kind of distribution after filtering with the condition $F-E^{homo} < 0.15t$.  }
    
    \label{fig:histo2}
\end{figure}

On the other hand, Fig.\ref{fig:histo2} illustrates the distribution of the density norm 
$|n - n^{\mathrm{homo}} |$ for the original dataset 
(panel~(a)) and for the filtered dataset satisfying the condition $F - E_{\mathrm{homo}} < 0.15t$ (panel~(b)).

In the original dataset, the distribution spans a broad range of 
density deviations from the homogeneous configuration, reflecting  
the large variety of disorder-induced ground states generated by 
the random external potentials. After applying the F-based 
filter, the distribution becomes strongly concentrated around 
small values of $| n - n^{\mathrm{homo}} |$, indicating that the retained samples correspond to weakly perturbed density  profiles close to the homogeneous reference state.

\section{The learning procedure and the neural network}
\label{sect:networklearning}

In supervised learning, the objective is to approximate an unknown target
function that maps input features to observable quantities.
This approximation is represented by a parametrized model
\(
f_{\boldsymbol{\theta}}
\),
which defines a function
\begin{equation}
f_{\boldsymbol{\theta}} : \mathcal{X} \longrightarrow \mathcal{Y},
\end{equation}
where $\mathcal{X}$ denotes the space of input features and $\mathcal{Y}$ the
space of target values.
The vector $\boldsymbol{\theta}$ collects all adjustable parameters of the
model and determines a specific element within a chosen family of functions.

Importantly, $f_{\boldsymbol{\theta}}$ is not restricted to neural networks.
Rather, it denotes a general hypothesis class whose form is fixed a priori,
while its parameters are inferred from data.
Common examples include linear and polynomial regression models, kernel-based
methods, decision trees, Gaussian processes, and neural networks.
In all cases, learning corresponds to selecting the parameters
$\boldsymbol{\theta}$ such that the model provides an accurate approximation to
the underlying input--output relation.

Given a dataset of input--output pairs
\(
\{(\mathbf{X}_j, y_j)\}_{j=1}^{N}
\),
the optimal parameters are determined by minimizing an empirical risk functional,
\begin{equation}
\boldsymbol{\theta}^\ast
=
\arg\min_{\boldsymbol{\theta}}
\frac{1}{N}
\sum_{j=1}^{N}
\mathcal{L}\!\left(
f_{\boldsymbol{\theta}}(\mathbf{X}_j),\, y_j
\right),
\end{equation}
where $\mathcal{L}$ is a loss function that quantifies the discrepancy between the
model prediction and the true target value.
This formulation is independent of the specific functional form of
$f_{\boldsymbol{\theta}}$ and applies uniformly to all parametrized learning
models.

In the present context, $f_{\boldsymbol{\theta}}$ represents a parametrized
approximation to the universal density functional, mapping local electronic
densities to the corresponding functional value.
Although neural networks are employed in this work due to their expressive
capacity and scalability, the formulation in terms of
$f_{\boldsymbol{\theta}}$ is completely general and would equally apply to other
learning models capable of representing nonlocal functionals.

The starting point of the machine-learning workflow for this work is the dataset, which is composed by the features $\{n_{i}\}$ obtained by exact diagonalization of the Hamiltonian corresponding to many independent realizations of the external random potential $\{v_{i}\}$. The data is then partitioned into exclusive subsets, namely, training, validation and test. So, in this context, the filtered dataset containing $N_{\mathrm{filt}}$ samples was partitioned into training, validation, and test subsets in the ratio $2:1:1$. The construction must guarantee the absence of data leakage. Data leakage occurs whenever the statistical independence between the training, validation, and test sets is violated, either directly or indirectly.

 The training set is used to determine the optimal parameters $\boldsymbol{\theta}^{*}$ of the model by minimizing a suitable loss function. Let $\{(\mathbf{x}_i, y_i)\}_{i=1}^{N_{\mathrm{tr}}}$ denote the training dataset, where $\mathbf{x}_i \in \mathbb{R}^{L}$ represents the input features (local electronic densities on a lattice of $L$ sites) and $y_i = \Delta F_i = F_i - E^{\mathrm{homo}}$ is the target value, since the filtered dataset contains values of the functional $F$ close to  $E^{\mathrm{homo}}$.

The neural network implements a parametrized function
$f_{\boldsymbol{\theta}} : \mathbb{R}^{L} \rightarrow \mathbb{R}$, and its
parameters $\boldsymbol{\theta}$ (weights and biases) are determined by minimizing the mean squared
error (MSE) loss function over the training set. This is performed iteratively over multiple epochs, where each epoch constitutes one full pass through the entire training dataset. During a training epoch, a mini-batch is a fixed-size subset \(B\) drawn from the complete training set containing \(N_{tr}\) samples. Each epoch consists of $
M = \left\lceil \frac{N_{tr}}{B} \right\rceil$ mini-batches, processed sequentially. Therefore, a mini-batch uses the parameter values that resulted from the previous one. In each iteration, the model performs a forward pass, computes the loss, and executes backpropagation only on the \(B\) samples of the current mini-batch, followed by a parameter update.

Let \(\mathcal{B}^{(t)} = \{(x_i, y_i)\}_{i=1}^B\) be a mini-batch of size \(B\) 
sampled from the training set at iteration \(t\). 
The loss computed on this mini-batch is:

\begin{equation}
\mathcal{L}_{\text{MSE}}^{(t)}(\bm{\theta}) = \frac{1}{B} \sum_{(x_i, y_i) \in \mathcal{B}^{(t)}} [f_{\bm{\theta}}(x_i) - y_i]^2.
\end{equation}

Parameters are updated after each mini-batch:
\begin{equation}
\bm{\theta}^{(t+1)} = \bm{\theta}^{(t)} - \eta \nabla_{\bm{\theta}} \mathcal{L}_{\text{MSE}}^{(t)}(\bm{\theta^{(t)}}),
\end{equation}
where \(\eta\) is the learning rate. 
A complete epoch involves processing \(\lceil N_{\text{tr}}/B \rceil\) mini-batches.

At the end of each training epoch, the generalization performance of the model
is assessed on the validation dataset
$\{(\mathbf{x}_j^{\mathrm{val}}, y_j^{\mathrm{val}})\}_{j=1}^{N_{\mathrm{val}}}$,
which is statistically independent from the training set. The validation error
is computed as the mean squared error
\begin{equation}
\mathcal{L}_{\mathrm{MSE}}^{\mathrm{val}}(\boldsymbol{\theta})
=
\frac{1}{N_{\mathrm{val}}}
\sum_{j=1}^{N_{\mathrm{val}}}
\left[
f_{\boldsymbol{\theta}}(\mathbf{x}_j^{\mathrm{val}}) -
y_j^{\mathrm{val}}
\right]^2,
\label{eq:mse_val}
\end{equation}
where the network parameters $\boldsymbol{\theta}$ are kept fixed and no
gradient updates are performed.

The validation error $\mathcal{L}_{\mathrm{MSE}}^{\mathrm{val}}$ is used solely
to control the training dynamics, such as early stopping and learning--rate scheduling, and does not enter the parameter optimization directly. Without this periodic validation, there would be no reliable signal to determine when to stop training, which model version performs best on unseen data, whether the model is memorizing training examples rather than learning generalizable patterns.

Finally, the test set is intended to provide a final and unbiased estimate of the generalization performance of the trained model. In this final step, the Mean Absolute Error (MAE) is a metric providing the average magnitude of prediction errors on the test set. It is computed once, after all training and validation Its formula is:

\begin{equation}
\text{MAE} = \frac{1}{N_{test}} \sum_{i=1}^{N_{test}} |y_i - \hat{y}_i|
\end{equation}

where $N_{test}$ is the number of test samples, \(y_i\) the true value, and \(\hat{y}_i = f_{\bm{\theta}^{*}}(\mathbf{x_{i}})\) the predicted value. It means that on average, each prediction is off by certain units of the target variable.

Thereforre, this strict separation of roles (training, validation and test) is essential to avoid data leakage and to ensure the statistical reliability of the reported results.

The universal density functional $F[n]$ is approximated using a
one-dimensional convolutional neural network (CNN) specifically designed
to respect the lattice structure and periodic boundary conditions of the
one--dimensional Hubbard model.
A schematic representation of the network architecture is shown in
Fig.~\ref{fig:cnn_vertical_architecture}.

The input to the network is the local density vector
$\mathbf{n}=(n_1,\dots,n_L)$ defined on a lattice of $L$ sites. To enforce periodic boundary conditions at the level of the convolutional
operation, the input vector is extended following the prescription of Ref.~\cite{nelson2019} by appending the first $k-1$ density components to the end of the vector,
\begin{equation}
\mathbf{n}_{\mathrm{ext}} =
(n_1,\dots,n_L,n_1,\dots,n_{k-1}),
\end{equation}
resulting in an input of length $L+k-1$.

The size of the kernel window is fixed to $k=3$ \cite{nelson2019}, so that the extended input has length $L+2$. This unilateral periodic wrapping ensures that convolutional windows can probe local density neighborhoods crossing the physical boundary of the lattice.

In order to be processed by a one--dimensional convolutional layer, the
input is reshaped into a tensor of shape $(L+k-1)\times 1$, where the
additional dimension corresponds to a single input channel.

The reshaped input is then processed by a single convolutional layer
consisting of $8$ independent filters with kernel size $k=3$ and unit
stride.
Each filter is defined by a set of trainable weights
$(w_1,w_2,\dots,w_k)$ and operates through a sliding window
(\emph{kernel window}) that scans contiguous segments of the density
sequence.
At each convolution step, the kernel acts on $k$ adjacent density values
(e.g., $(n_i,n_{i+1},n_{i+2})$ for $k=3$), producing one output value
associated with the corresponding window position.

By extending the density vector through periodic wrapping, the sliding
windows include neighborhoods that cross the boundary of the lattice.
As a consequence, every lattice site $n_i$ appears as part of a kernel
window together with its true physical neighbors, including the periodic
neighbors at the boundary (e.g., $(n_{L-1},n_L,n_1)$ and $(n_L,n_1,n_2)$).
Each filter is therefore applied uniformly across all spatial positions, ensuring translational invariance and enforcing periodic boundary
conditions at the level of the convolutional features.
The convolutional kernels are thus forced to learn patterns compatible with the cyclic topology of the system, rather than spurious edge effects (See Fig.\ref{fig:filtro}).

As illustrated in Fig.\ref{fig:filtro}, the convolutional layer uses an activation function (AF), which in most cases is a rectified linear unit (ReLU) activation function.

The output of the convolutional layer, consisting of $\ell$ spatial positions and $8$ channels, is flattened into a one-dimensional feature vector of dimension $8\ell$. This representation is then passed through two fully connected layers, each containing $128$ hidden units with the selected activation function. Finally, a linear output layer produces a single scalar value $F_{\mathrm{ML}}$, representing the network prediction for the exact density functional $F[n]$.

The network parameters are optimized by minimizing the mean-squared error (MSE) loss between the predicted functional value $F_{\mathrm{ML}}$ and the exact value. As mentioned before, the exact ground state values of $F$ are obtained from exact diagonalization of the matrix Hamiltonian.\\

Training is performed using the Adam optimizer with a learning rate of
$10^{-3}$ \cite{Ruder2016}. No dropout or $L_2$ regularization is employed, as the model capacity is kept intentionally small in order to emphasize physically meaningful local correlations rather than overparameterization \cite{Srivastava2014}.
\subsection{Supervised Learning of the functional $F[\mathbf{n}]$ as the only target}

In order to assess which physical input carries the relevant information for predicting the universal functional $F$, as a mean to test the universal character of the functional $F$, we performed an ablation study comparing models trained with different input features: (i) the charge density only ($\mathbf{n}$), (ii) the external potential only ($\mathbf{v}$), and (iii) the combined input $(\mathbf{n},\mathbf{v})$. Each model was trained using the identical  network architecture and hyperparameters described in section \ref{sect:networklearning}, differing solely in the input representation.

To evaluate generalization performance and detect possible overfitting, we monitored the mean absolute error (MAE) of the predicted functional $\Delta F = F - E^{\mathrm{homo}}$ on a held-out validation set as a function of the training epoch, due to the fact that the filtered dataset contains samples in which $F$ is close to $E^{homo}$. The evolution of the validation MAE provides a direct measure of how efficiently each input representation allows the model to learn the mapping between the physical configuration (obtained by exact diagonalization of the Hamiltonian) and the corresponding functional value, independently of training-set memorization.

Figure~\ref{fig:mae_val} shows the evolution of the validation mean absolute error (MAE) for the predicted functional $\Delta F$ as a function of the training epoch, for models trained using three different input representations: density only ($\mathbf{n}$), potential only ($\mathbf{v}$), and the combined input $(\mathbf{n},\mathbf{v})$. All models exhibit a rapid initial decrease in validation error followed by convergence to a stable plateau, indicating effective learning without significant overfitting due to the use of early stopping.

It can be observed that the model trained solely on the density achieves the lowest validation MAE throughout most of the training process, closely followed by the model using both $(\mathbf{n},\mathbf{v})$ as input. In contrast, the model trained exclusively on the external potential consistently attains higher validation error and converges more slowly. This behavior confirms that the relevant predictive information for estimating the functional $F$ is predominantly encoded in the charge density, while the inclusion of the external potential does not lead to a substantial improvement in generalization performance. The comparatively slightly poorer performance of the $v$-only model supports the interpretation that the dependence of the functional on the external potential is mediated indirectly through the density, in agreement with the Hohenberg-Kohn theorem.

In order to improve the study of the learning procedure of the functional $F$ and its dependence of the density we must consider that the one-dimensional Hubbard model considered here is invariant under spatial symmetry operations including discrete translations and mirror reflections. In particular, given a charge density profile $n = \{n_i\}_{i=1}^{L}$ defined on a ring of length $L$, the translated density is defined as
\begin{equation}
[T_s(n)]_i = n_{i+s \; (\mathrm{mod}\, L)},
\end{equation}
for an integer shift $s$, while the mirror-reflected density is given by
\begin{equation}
[M(n)]_i = n_{L+1-i}.
\end{equation}
Due to the invariance of the Hamiltonian under these operations, the corresponding ground-state energies and universal functional values remain unchanged, i.e.,
\begin{equation}
F[n] = F[T_s(n)] = F[M(n)].
\end{equation}

Following the procedure proposed in Ref.\cite{nelson2019}, we augmented the training dataset by applying these symmetry operations to each density configuration, thereby generating additional samples belonging to the same equivalence class under the symmetry group of the lattice. Since these transformed densities describe physically equivalent systems, they were assigned the same target functional value $F[n]$. This augmentation procedure effectively constrains the learned functional to be invariant under the spatial symmetries of the system and improves the representation of the physically relevant manifold of densities in the input space.

Figure~\ref{fig:pred1} shows the prediction results obtained from a model trained without symmetry-augmented data. Panel (a) presents the comparison between the exact functional values $F_{\mathrm{exact}}$ and the corresponding predictions $F_{\mathrm{ML}}$, while panel (b) displays the residuals $F_{\mathrm{exact}} - F_{\mathrm{ML}}$ as a function of $F_{\mathrm{ML}}$. Although a reasonable agreement is observed, the residuals exhibit a noticeable spread, indicating limited generalization performance. Furthermore, in the absence of symmetry augmentation, the model trained using density-only inputs achieves a test error of $\mathrm{MAE}(F) \approx 1.7\times10^{-3}$.

In contrast, Figure~\ref{fig:pred2} presents the corresponding results obtained from a model trained with symmetry-augmented densities. As shown in panel (a), the predicted values exhibit an almost perfect linear correlation with the exact functional values. Moreover, panel (b) reveals a substantial reduction in the magnitude and dispersion of the residuals across the entire range of predicted values.  When symmetry-augmented densities are incorporated into the training and validation sets, this error is reduced to $\mathrm{MAE}(F) \approx 2.93\times10^{-4}$ on the same unaugmented test set, corresponding to an improvement by approximately a factor of six.

\section{Prediction of the potentials}

 In Ref.\cite{nelson2019} the prediction of the external potential appeared to be performed through a separate supervised learning task. After training the network to reproduce the universal functional $F[n]$, the same CNN architecture is retrained using the density as input and the corresponding external potential $\{v_{i}\}$ as output, thus learning the mapping $\mathbf{n} \mapsto \mathbf{v}$ directly from data.

In contrast, we use a different approach in which the potential is not learned as an independent supervised output. Instead, it is obtained from the functional derivative of the learned universal functional, though $F[\mathbf{n}]$ is the only output of the CNN. 
In the Hohenberg--Kohn formalism, the ground-state density of an interacting many-body system in the presence of an external potential  is obtained by minimizing the energy functional given in Eq.~(\ref{eq:identidade}) under the particle-number constraint. So  
\begin{equation}
E_v[n] = F[n] + 2\sum_i v_i n_i,
\end{equation}
among all densities satisfying the particle-number constraint
$\sum_i n_i = N_{\uparrow}$.

Introducing a Lagrange multiplier associated with this constraint, one considers the auxiliary functional
\begin{equation}
\Phi[n] =
F[n] + 2\sum_i v_i n_i
-\mu\!\left(\sum_i n_i - N_{\uparrow}\right).
\end{equation}
Stationarity with respect to admissible density variations
$\delta n_i$ satisfying $\sum_i \delta n_i = 0$ yields the
Euler-Lagrange condition
\begin{equation}
\frac{\partial F[n]}{\partial n_i} + 2v_i = \mu,
\qquad \text{evaluated at } n=n^\star.
\end{equation}
Therefore, the external potential is related to the functional
derivative of $F[n]$ by
\begin{equation}
v_i[n] = -\frac{1}{2}\frac{\partial F[n]}{\partial n_i} + \frac{1}{2}\mu.
\label{eq:derivadaf}
\end{equation}

The value of  $\mu$ could be absorbed into the gauge freedom of the external potential in fixed-particle-number sectors. In practice, this gauge freedom is removed by subtracting the spatial mean of the potential, allowing a direct comparison between the reconstructed and exact potentials.\\


Nevertheless, training a machine-learning model to reproduce
$F[n]$ alone does not guarantee that its gradient correctly represents the physical potential. Indeed, minimizing a loss function based solely on functional values enforces agreement between $F_{\mathrm{ML}}[n]$ and the exact functional on the sampled density manifold, but does not constrain their
functional derivatives. As a consequence, the learned model may differ from the exact
functional by a residual functional error induced by the
trained parameters of the neural network.
Denoting by $F_\theta[n]$ the parametric approximation of the
universal functional, the learned functional after training
could be written as
\begin{equation}
F_{\mathrm{ML}}[n]
=
F_{\theta^\star}[n]
=
F[n]
+
\varepsilon_{\theta^\star}[n],
\label{eq:F_decomposition}
\end{equation}
where
\begin{equation}
\varepsilon_\theta[n]
=
F_\theta[n]
-
F[n].
\end{equation}
The training procedure minimizes the loss
$\mathcal{L}_F(\theta)$ and therefore constrains the magnitude of
$\varepsilon_{\theta^\star}[n]$ over the training manifold, typically
leading to
\begin{equation}
|\varepsilon_{\theta^\star}[n]|
\ll
|F[n]|.
\end{equation}
However, since the loss penalizes only discrepancies
in the functional values, it does not constrain the functional derivative of the residual term. Taking the functional derivative of
Eq.~(\ref{eq:F_decomposition}) yields
\begin{equation}
\frac{\partial F_{\mathrm{ML}}[n]}{\partial n_i}
=
\frac{\partial F[n]}{\partial n_i}
+
\frac{\partial \varepsilon_{\theta^\star}[n]}{\partial n_i}.
\end{equation}
The reconstructed potential obtained from the learned functional is
\begin{equation}
v_i^{\mathrm{ML}}[n]
=
-\frac{1}{2}\frac{\partial F_{\mathrm{ML}}[n]}{\partial n_i}
+
\frac{1}{2}\tilde\mu,
\label{eq:obtervml}
\end{equation}
which becomes
\begin{equation}
v_i^{\mathrm{ML}}[n]
=
v_i[n]
+
\frac{1}{2}(\tilde\mu-\mu)
-
\frac{1}{2}\frac{\partial \varepsilon_{\theta^\star}[n]}{\partial n_i}.
\label{eq:vML_eps}
\end{equation}
Thus, the predicted potential differs from the physical one
not through a global rescaling of the functional, but through
an additional contribution arising from the gradient of the
residual functional error.

Since the external potential is defined only up to an additive
constant, we fix its gauge by subtracting the spatial mean,
\begin{equation}
v_i^{\mathrm{exact(gauge)}}
=
v_i
-
\bar v,
\qquad
\bar v=\frac{1}{L}\sum_j v_j .
\end{equation}
After gauge fixing, Eq.~(\ref{eq:vML_eps}) implies
\begin{equation}
v_i^{\mathrm{ML(gauge)}}
\approx
v_i^{\mathrm{exact(gauge)}}
-
\frac{1}{2}\frac{\partial \varepsilon_{\theta^\star}[n]}{\partial n_i},
\end{equation}
showing that inaccuracies in the reconstructed potential are
governed by the gradient of the residual functional error. To illustrate this, Figure~\ref{fig:pred_pot1} shows the comparison between the gauge-fixed reconstructed potentials $v_i^{\mathrm{ML}(gauge)}$ and the
corresponding exact gauge-fixed external potentials $v_i^{\mathrm{exact}(gauge)}$
for representative samples from the test dataset. Despite the accurate prediction of the functional values $F[n]$, as in Fig.{\ref{fig:pred2}}, the reconstructed
potentials exhibit noticeable deviations from the exact ones even after gauge fixing. This relevant deviations are due to the size of the norm of the gradient of the residual functional $\nabla \varepsilon_{\theta^{\star}}[\mathbf{n}]$.  In Fig.\ref{fig:gradiente} is shown the histogram of  $|\nabla \varepsilon_{\theta^{\star}}[\mathbf{n}]|$. Note that its mean value is approximately $0.69$, which is not small. Thus, this is the source of the erro in predicting the potential $\mathbf{v}$. \\


Therefore, minimizing a loss function based solely on the discrepancy between the learned and exact functional values,
\begin{equation}
\mathcal{L}_F
=
\left\langle
\left(F_{\mathrm{ML}}[\mathbf{n}] - F_{\mathrm{exact}}[\mathbf{n}]\right)^2
\right\rangle,
\label{eq:LF_only}
\end{equation}
is insufficient to ensure the physical consistency of the learned functional in a variational context. To address this issue, it is necessary to introduce an additional term in the loss function that provides a condition relating the functional derivative of $F[\mathbf{n}]$ to the external potential. \\\\
When the network is trained directly on the physical functional values,
global offsets and rescalings in $F[\mathbf n]$ may alter its functional
derivative. As a consequence, a model that accurately reproduces the values
of $F[\mathbf n]$ may still yield inconsistent potentials when evaluated
through automatic differentiation.\\

To improve numerical conditioning during training, it is convenient to predict
a normalized version of the functional defined as
\begin{equation}
F_n[\mathbf n]
=
\frac{F_{\mathrm{exact}}[\mathbf n]-F_{\mathrm{mean}}}{F_{\mathrm{std}}},
\label{eq:Fn}
\end{equation}
where $F_{\mathrm{mean}}$ and $F_{\mathrm{std}}$ denote the mean and standard
deviation of the exact functional values over the training set, respectively.
Target normalization remains advisable even in the presence of adaptive
weighted losses, since unaddressed magnitude discrepancies between the
function and gradient terms may otherwise disrupt training stability and
convergence.\\

The CNN receives the density $\mathbf n$ as input and outputs a scalar
quantity $F_{\mathrm{net}}[\mathbf n]$, which represents a normalized
approximation to the target functional $F_n[\mathbf n]$.
The functional is learned directly in its normalized representation. \\

The total loss function suited to learn the functional and its gradient is defined as
\begin{equation}
\mathcal L_{\mathrm{tot}}
=
\sum_{l \in \{F,EL\}}
\left(
e^{-s_l}\,\mathcal L_l + s_l
\right),
\label{eq:Ltot}
\end{equation}
where the individual loss components are given by
\begin{equation}
\mathcal L_F =
\left\langle
\left(
F_{\mathrm{net}}[\mathbf n]
-
F_n[\mathbf n]
\right)^2
\right\rangle,
\end{equation}
\begin{equation}
\mathcal L_{EL} =
\left\langle
\sum_i
\left(
\frac{\partial F_{\mathrm{ML}}[\mathbf n]}{\partial n_i}
+
2v_i^{\mathrm{exact(gauge)}}[\mathbf n]
\right)^2
\right\rangle.
\label{eq:L_EL}
\end{equation}
Here, the quantities $s_l$ are trainable log-variance parameters that
dynamically regulate the relative contribution of each loss term during
optimization. Each term $\mathcal L_l$ is weighted by a factor $e^{-s_l}$,
allowing the optimization procedure to adaptively balance the contributions
of the functional loss and the Euler--Lagrange residual without requiring
manual tuning of fixed hyperparameters.\\

The learned functional in physical $t=1$ units is reconstructed as
\begin{equation}
F_{\mathrm{ML}}[\mathbf n]
=
F_{\mathrm{mean}}
+
F_{\mathrm{std}}\,
F_{\mathrm{net}}[\mathbf n].
\label{eq:Fphys_rec}
\end{equation}

During training, the functional derivative entering
Eq.~(\ref{eq:L_EL}) is obtained from the normalized
representation of the learned functional through
the chain rule. Since the physical functional is
reconstructed according to Eq.~(\ref{eq:Fphys_rec}),
its functional derivative reads
\begin{equation}
\frac{\partial F_{\mathrm{ML}}[\mathbf n]}{\partial n_i}
=
F_{\mathrm{std}}\,
\frac{\partial F_{\mathrm{net}}[\mathbf n]}{\partial n_i}.
\label{eq:dF_chain}
\end{equation}

In practice, the derivative
$\partial F_{\mathrm{net}}/\partial n_i$
is computed by automatic differentiation
with respect to the network output, and
subsequently rescaled by $F_{\mathrm{std}}$
to recover the derivative of the physical functional.\\
This automatic differentiation is implemented in a TensorFlow framework. In this approach, the network output $F_{\mathrm{net}}[\mathbf n]$
is represented as a composition of elementary
differentiable operations acting on the input
density vector $\mathbf n$. During the forward pass, the computational graph associated with this composition is constructed explicitly.\\

The machine-learned potentials are then obtained as
\begin{equation}
v_i^{\mathrm{ML}(gauge)}[\mathbf n]
=
-
\frac{1}{2}\frac{\partial F_{\mathrm{ML}}[\mathbf n]}{\partial n_i}.
\label{eq:rec_pot}
\end{equation}

\begin{figure}[ht]
    \centering
    \includegraphics[width=1.0\textwidth]{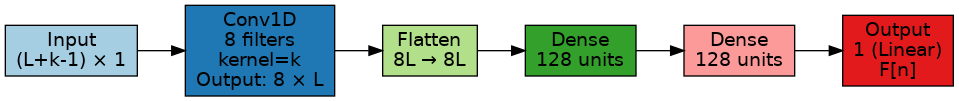}
\caption{Schematic representation of the convolutional neural
network architecture for learning the functional $F[\{n_{i}\}]$.
The input density vector is extended by a unilateral periodic wrap, as $(n_{1},...,n_{L},n_{1},..,n_{k-1})$, consistent with the dataset generation procedure and with
Ref.~\cite{nelson2019}, enabling a one-dimensional convolution with kernel size $k=3$ to probe local neighborhoods crossing the lattice boundary. The kernel size also defines the spatial range of non-locality captured by the learned functional.}
\label{fig:cnn_vertical_architecture}
\end{figure}

\begin{figure}[ht]
    \centering
    \includegraphics[width=0.5\textwidth]{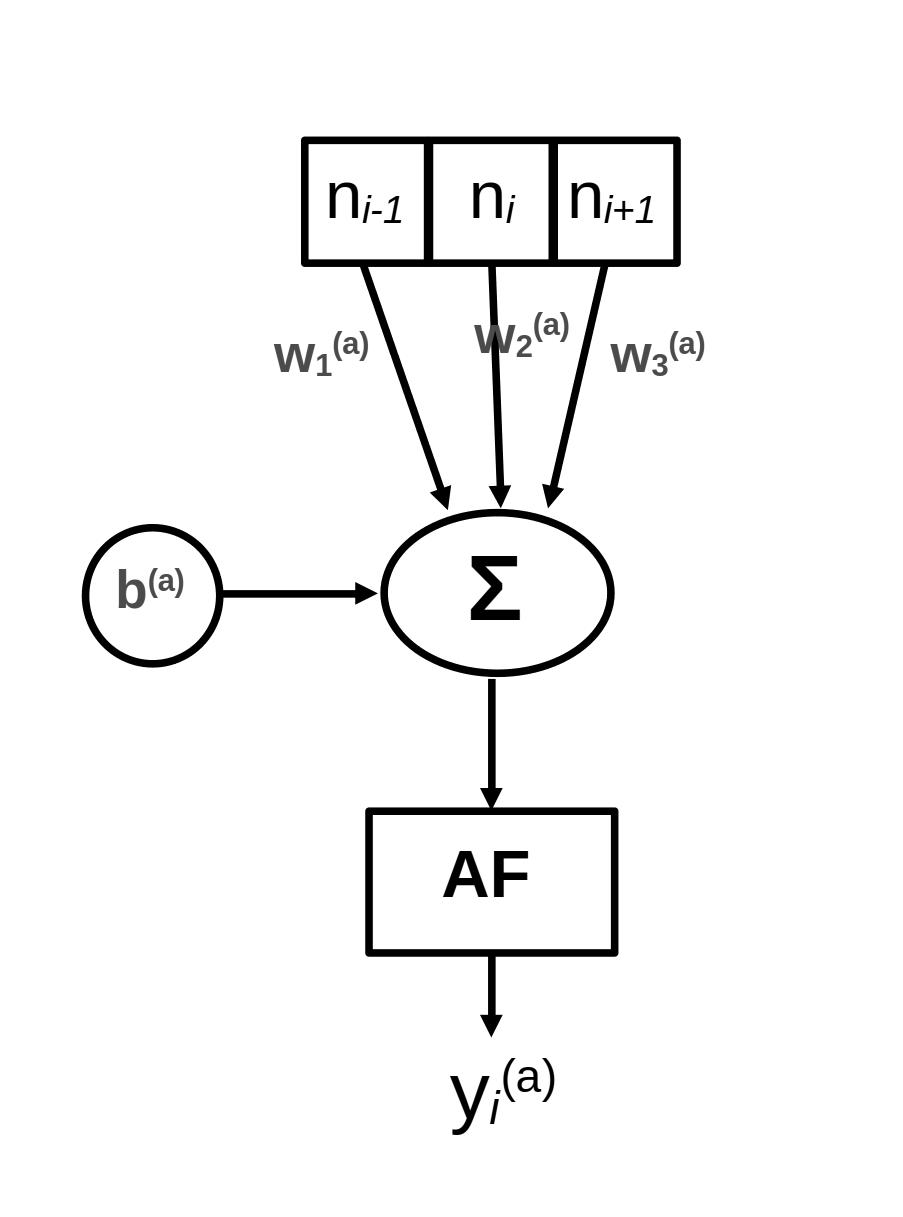}
    \caption{Illustration of the operation that the filter of index $a$ of the convolutional layer applies to the kernel window (size k=3) corresponding to the site $i$. The output is produced by the activation function (AF), so that $y_{i}^{a} = \mathbf{AF}(w_{1}^{a}n_{i-1}+w_{2}^{(a)}n_{i}+w_{3}^{(a)}n_{i+1}+b^{(a)})$, where $w_{1}^{(a)},w_{2}^{(a)},w_{3}^{(a)}$ are the weights of filter $a$ and $b^{(a)}$ its bias. A filter applies this operation for each site of the lattice producing 8 output values. There are $L=8$ filters, so $a=1,...,L$. }
    \label{fig:filtro}
\end{figure}

\begin{figure}[ht]
    \centering
    \includegraphics[width=0.8\textwidth]{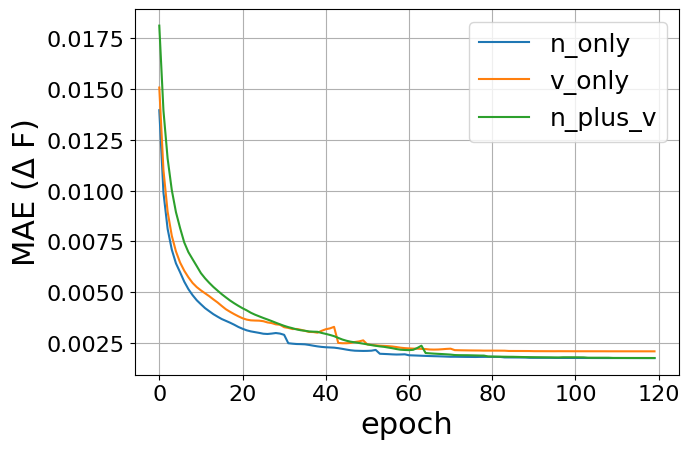}
\caption{Validation MAE of the predicted functional $\Delta F$ as a function of training epoch for models trained with density-only ($\mathbf{n}$), potential-only ($\mathbf{v}$), and combined ($\mathbf{n},\mathbf{v}$) inputs. The density-only model achieves the lowest validation error, indicating that the relevant predictive information is primarily encoded in the charge density.}
\label{fig:mae_val}
\end{figure}

\begin{figure}[h]
    \centering
    
    \begin{subfigure}{0.45\textwidth}
        \centering
        \includegraphics[width=\linewidth]{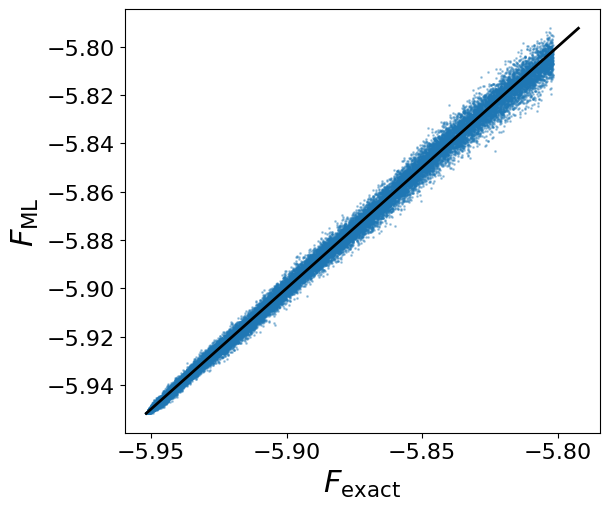}
        \caption{}
        \label{fig:a}
    \end{subfigure}
    \begin{subfigure}{0.45\textwidth}
        \centering
        \includegraphics[width=\linewidth]{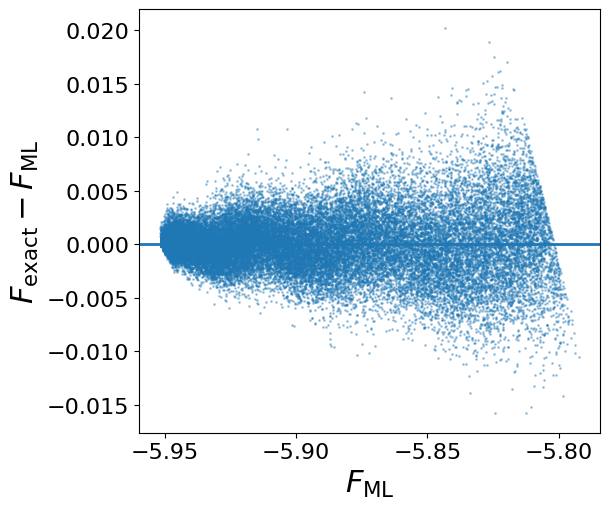}
        \caption{}
        \label{fig:b}
    \end{subfigure}
    \caption{Prediction of the universal functional $F$ on the test set using density-only inputs without symmetry-augmented data: (a) $F_{\mathrm{ML}}$ versus $F_{\mathrm{exact}}$, and (b) residuals $F_{\mathrm{exact}}-F_{\mathrm{ML}}$ as a function of $F_{\mathrm{ML}}$. }
    
    \label{fig:pred1}
\end{figure}

\begin{figure}[h]
    \centering
    
    \begin{subfigure}{0.45\textwidth}
        \centering
        \includegraphics[width=\linewidth]{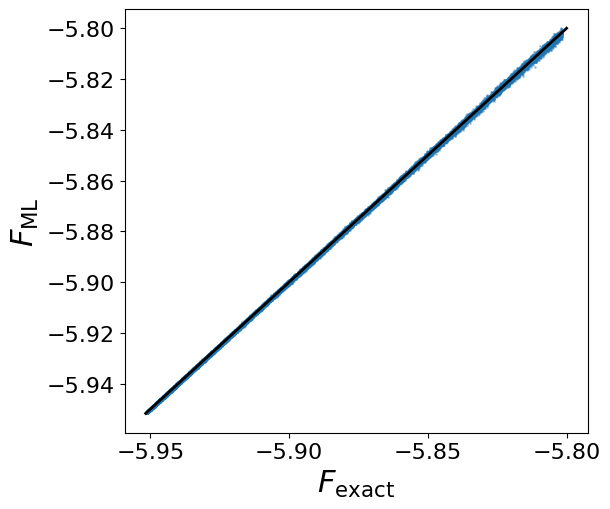}
        \caption{}
        \label{fig:a}
    \end{subfigure}
    \begin{subfigure}{0.45\textwidth}
        \centering
        \includegraphics[width=\linewidth]{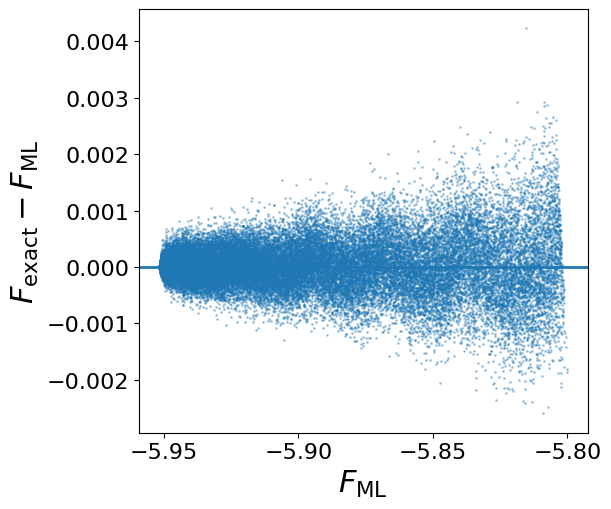}
        \caption{}
        \label{fig:b}
    \end{subfigure}
    \caption{Prediction of the universal functional $F$ on the test set using density-only inputs with symmetry-augmented densities (translations and mirror reflections): (a) $F_{\mathrm{ML}}$ versus $F_{\mathrm{exact}}$, and (b) corresponding residuals. The inclusion of symmetry-related configurations leads to a significant reduction in prediction error on the unaugmented test set.} 
    
    \label{fig:pred2}
\end{figure}

\begin{figure}[ht]
    \centering
    \includegraphics[width=0.8\textwidth]{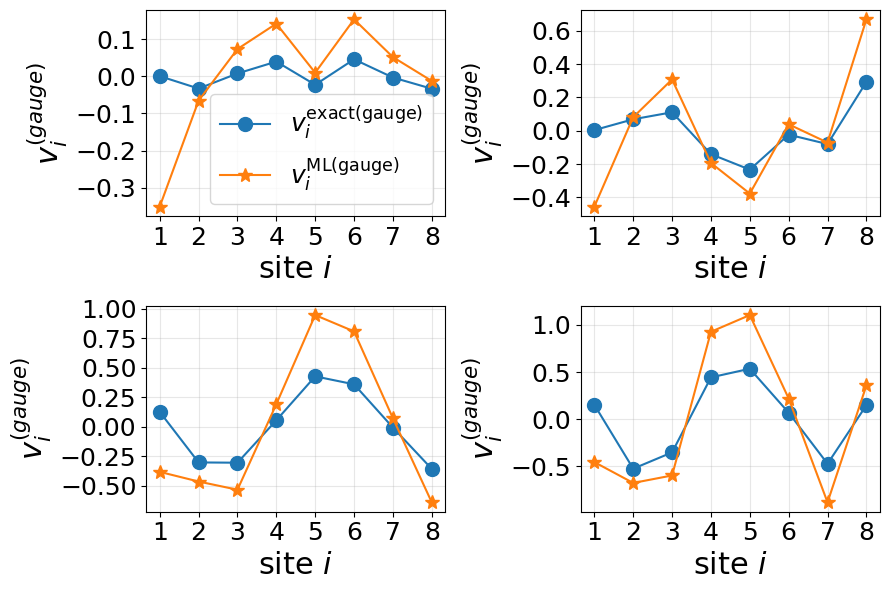}
\caption{Gauge-fixed external potentials reconstructed from the learned
functional via $v_i^{\mathrm{ML}} = -\partial F_{\mathrm{ML}}/\partial n_i$,
compared to the exact gauge-fixed potentials for representative test samples.
Noticeable discrepancies remain after gauge fixing, reflecting relevant errors in the
functional derivative when training the loss function in Eq.(\ref{eq:LF_only}). }
\label{fig:pred_pot1}
\end{figure}

\begin{figure}[ht]
    \centering
    \includegraphics[width=0.8\textwidth]{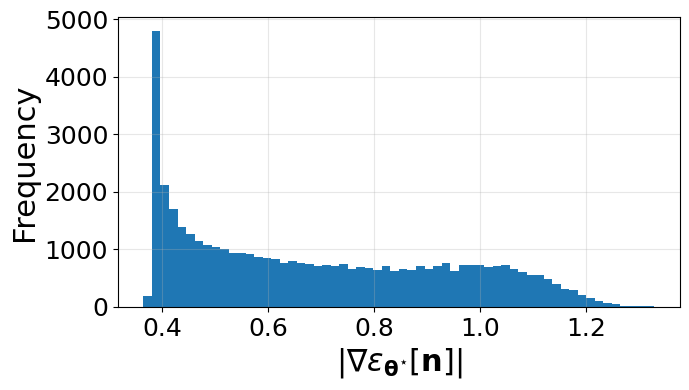}
\caption{Histogram of the norm of the residual functional gradient evaluated over the test set,
where $\varepsilon_{\theta^\star}[n] = F_{\mathrm{ML}}[n] - F[n]$.
The distribution quantifies the variational error of the learned functional
and its impact on the reconstructed external potentials.}
\label{fig:gradiente}
\end{figure}

We implemented this training procedure in a CNN with the
same configuration shown in Fig.~\ref{fig:cnn_vertical_architecture},
with the exception of the activation function. In this
context we employed the softplus activation function,
which constitutes a smooth approximation to the rectified
linear unit (ReLU). Unlike ReLU, whose derivative is
discontinuous at the origin, the softplus function is
continuously differentiable, ensuring that the learned
functional belongs to a differentiability class compatible
with the variational Euler--Lagrange condition. This
smoothness prevents the appearance of spurious
discontinuities in the functional derivative obtained
via automatic differentiation implemented in TensorFlow,
leading to a more stable and physically consistent
reconstruction of the external potential.\\

\begin{figure}[ht]
    \centering
    \includegraphics[width=0.6\textwidth]{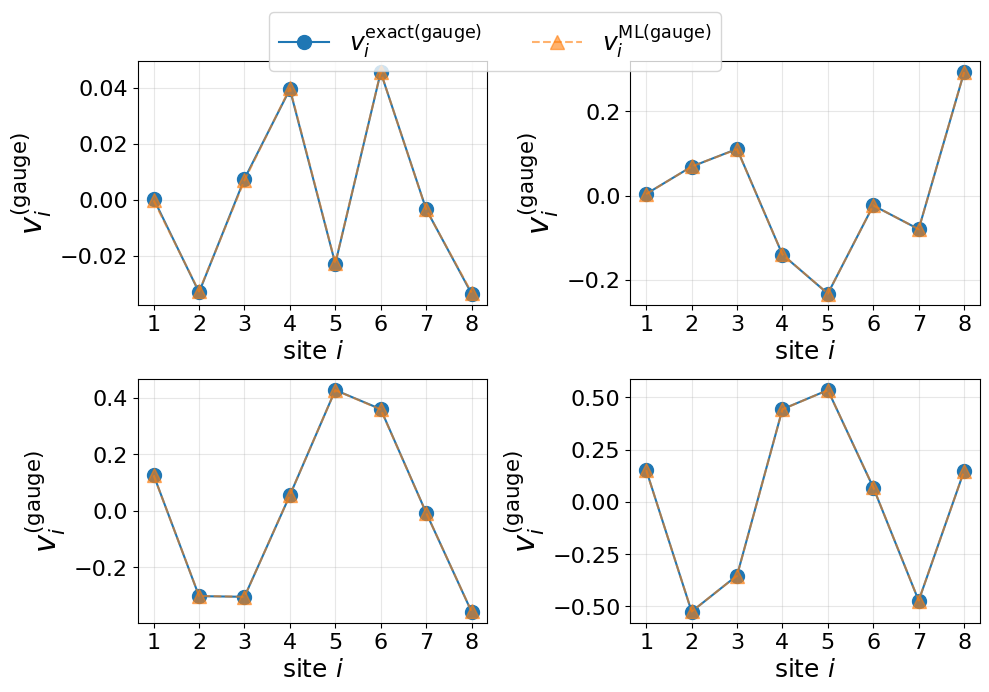}
\caption{Gauge-fixed external potentials reconstructed from the learned
functional via $v_i^{\mathrm{ML}{gauge)}} = -\partial F_{\mathrm{ML}}/\partial n_i$,
compared to the exact gauge-fixed potentials for representative test samples.}
\label{fig:pred_pot2}
\end{figure}
\begin{figure}[h]
    \centering
       \includegraphics[width=0.6\linewidth]{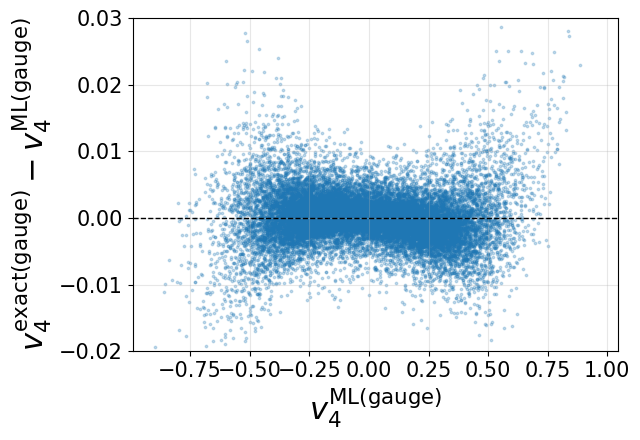}
       \caption{Residual of the  potential at site
$i=4$. The small mean absolute error is $1.7\times 10^{-3}$.} 
   \label{fig:final_predv4}
\end{figure}
\begin{figure}[h]
    \centering
        \begin{subfigure}{0.4\textwidth}
        \centering
        \includegraphics[width=\linewidth]{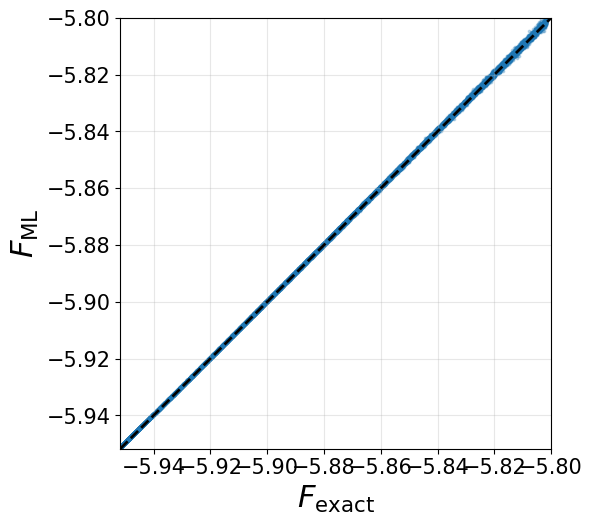}
        \caption{}
        \label{fig:a}
    \end{subfigure}
    \begin{subfigure}{0.5\textwidth}
        \centering
        \includegraphics[width=\linewidth]{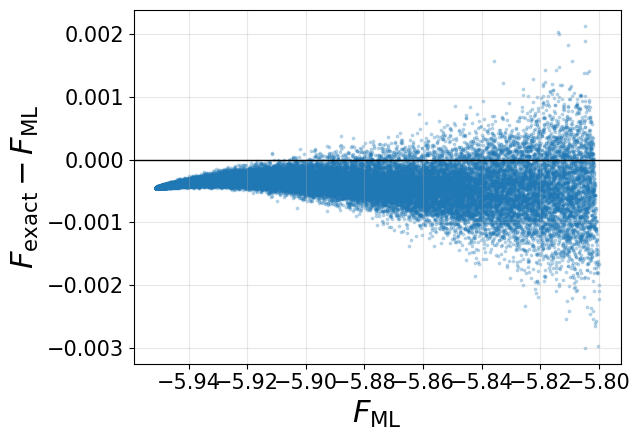}
        \caption{}
        \label{fig:b}
    \end{subfigure}
    \caption{ (a) Predicted functional $F_{ML}$ versus $F_{exact}$; (b) Residual between the learned functional $F_{\mathrm{ML}}$ and the exact functional. The  mean absolute error is $4\times10^{-4}$. } 
   \label{fig:final_pred}
\end{figure}

We implemented this training procedure in a CNN with the same configuration shown in Fig.\ref{fig:cnn_vertical_architecture} with the exception of the activation function. In this context we employed the softplus activation function, which constitutes a smooth approximation to the rectified linear unit (ReLU). Unlike ReLU, whose derivative is discontinuous at the origin, the softplus
function is continuously differentiable, ensuring that the learned functional
 belongs to a differentiability class compatible with the variational Euler-Lagrange condition. This smoothness prevents the appearance
of spurious discontinuities in the functional derivative obtained via automatic differentiation implemented in TensorFlow, leading to a more stable and physically consistent
reconstruction of the external potential.\\
The results in predicting the potentials directly from the learned gradient is shown in Fig.\ref{fig:pred_pot2}. There, the predicted potentials $v_{i}^{ML(gauge)}$ of four samples of the test set (the same used in Fig.\ref{fig:pred_pot1}) exhibit high agreement with the exact values $v_{i}^{exact(gauge)}$. With regard to this, the residual error of the predicted potential of site 4 (using the dataset) is shown in Fig.\ref{fig:final_predv4}. The degree of accuracy is the same as that exhibited in Fig.2 of Ref.\cite{nelson2019}.\\ 
In addition,  Fig.~\ref{fig:final_pred} shows resulting predicted functional. It can be observed that the the accuracy in the  predicted functional is just the same as that shown in Fig.\ref{fig:pred2}. However, in Fig.~\ref{fig:final_pred}(b) the residual is asymmetric. The asymmetry observed in the residual distribution can be attributed to a small global affine mismatch between the predicted and exact functional values. \\

In order to quantify any residual global mismatch between the predicted and exact functional values, we perform an affine calibration of the machine-learned functional using the validation set. Specifically, the parameters $a_{\mathrm{val}}$ and $b_{\mathrm{val}}$ are determined by solving the least-squares problem
\begin{equation}
(a_{\mathrm{val}}, b_{\mathrm{val}})
=
\arg\min_{a,b}
\left\langle
\left(
F_{\mathrm{exact}}[\mathbf{n}]
-
a\,F_{\mathrm{ML}}[\mathbf{n}]
-
b
\right)^2
\right\rangle_{\mathrm{val}},
\end{equation}
which corresponds to the best global affine mapping between the predicted and exact functional values on the validation set. The calibrated functional used for evaluation on the test set is then defined as
\begin{equation}
F_{\mathrm{ML}}^{\mathrm{cal}}[\mathbf{n}]
=
a_{\mathrm{val}}\,F_{\mathrm{ML}}[\mathbf{n}]
+
b_{\mathrm{val}}.
\end{equation}

 The validation-based calibration yields parameters $a_{\mathrm{val}}\simeq 0.99936 \approx 1$ and $b_{\mathrm{val}}\simeq -4.17\times10^{-3}$, indicating a residual offset that manifests as a systematic skew in the error distribution.
\section{Conclusions}

We have revisited the machine-learning construction of the universal density
functional for the one-dimensional Hubbard model in a site-dependent random
potential, using exact diagonalization data for a periodic chain with $L=8$ in
the $(2,2)$ sector. \\

By performing PCA in the joint feature space $(\mathbf n,\mathbf v)$, we found
that the dataset occupies a low-dimensional manifold consistent with the
physical map $\,\mathbf v \mapsto \mathbf n(\mathbf v)\,$. Importantly, the PCA
also isolates a direction almost entirely confined to the potential subspace,
corresponding to uniform shifts of $v_i$; this provides a clear data-driven
signature of the gauge freedom $v_i \to v_i + c$ in fixed-particle-number
sectors. \\
An energy-based filtering criterion,
$F-E^{\mathrm{homo}}<0.15t$, was shown to restrict the dataset to weakly
perturbed configurations, leading to a smoother and more monotonic dependence of
$F$ on the distance to the homogeneous density and improving the conditioning of
the supervised learning task. Symmetry augmentation through translations and mirror reflections
further improves generalization and yields near-exact predictions of $F[\mathbf
n]$ with a compact 1D CNN architecture that respects periodicity via unilateral \\
wrapping.

A central outcome of this work is to clarify that high accuracy in predicting
functional values does not automatically translate into accuracy of functional
derivatives. Since the reconstructed potential follows from
$v_i^{\mathrm{ML}}=-\partial F_{\mathrm{ML}}/\partial n_i$ (up to a constant),
errors in the residual functional can be amplified at the derivative level.
To mitigate this issue, we augmented the training objective with a variational
consistency term enforcing the Euler--Lagrange relation against the gauge-fixed
exact potentials, and we employed a simple affine calibration determined on a
validation set to understand global slope/offset effects in the functional values.
\\
\section*{Acknowledgments}
The authors acknowledge the financial support from the National Council for Scientific and Technological Development (CNPq) - Brazil.









\end{document}